\documentclass[12pt]{article}
\title{Hamiltonian and Brownian systems with long-range interactions: IV. General kinetic equations \\
from the quasilinear theory}

\usepackage{amsthm,amsmath,amssymb,a4wide,epsf}

cm \hsize=18 true cm \topmargin=-2cm \oddsidemargin=-0.5cm
\def\mb#1{\setbox0=\hbox{$#1$}\kern-.025em\copy0\kern-\wd0
\kern-0.05em\copy0\kern-\wd0\kern-.025em\raise.0233em\box0}

\begin{document}

\author{Pierre-Henri Chavanis}
\maketitle
\begin{center}
Laboratoire de Physique Th\'eorique (CNRS UMR 5152), \\
Universit\'e
Paul Sabatier,\\ 118, route de Narbonne, 31062 Toulouse Cedex 4, France\\
E-mail: {\it chavanis{@}irsamc.ups-tlse.fr\\
 }
%\date{}
\vspace{0.5cm}
\end{center}

\begin{abstract}

We develop the kinetic theory of Hamiltonian systems with weak
long-range interactions. Starting from the Klimontovich equation and
using a quasilinear theory, we obtain a general kinetic equation that
can be applied to spatially inhomogeneous systems and that takes into
account memory effects.  This equation is valid at order $1/N$ in a
proper thermodynamic limit and it coincides with the kinetic equation
obtained from the BBGKY hierarchy.  For $N\rightarrow +\infty$, it
reduces to the Vlasov equation describing collisionless systems. We
describe the process of phase mixing and violent relaxation leading to
the formation of a quasi stationary state (QSS) on the coarse-grained
scale. We interprete the physical nature of the QSS in relation to
Lynden-Bell's statistical theory and discuss the problem of incomplete
relaxation. In the second part of the paper, we consider the
relaxation of a test particle in a thermal bath. We derive a
Fokker-Planck equation by directly calculating the diffusion tensor
and the friction force from the Klimontovich equation. We give general
expressions of these quantities that are valid for possibly spatially
inhomogeneous systems with long correlation time. We show that the
diffusion and friction terms have a very similar structure given by a
sort of generalized Kubo formula. We also obtain non-markovian kinetic
equations that can be relevant when the auto-correlation function of
the force decreases slowly with time.  An interest of our approach is
to develop a formalism that remains in physical space (instead of
Fourier space) and that can deal with spatially inhomogeneous systems.

\end{abstract}
\eject

\section{Introduction}
\label{sec_introduction}

In the treatment of physical systems in nature, a fundamental
distinction must be made between systems for which the interaction
between particles is short-range or long-range. Systems with
short-range interactions have been studied for a very long time. They
are spatially homogeneous, the ordinary thermodynamic limit
($N\rightarrow +\infty$ with fixed $N/V$) applies, and the statistical
ensembles are equivalent for $N\rightarrow +\infty$. In recent years,
a growing number of physical systems with truly long-range
interactions have emerged and have been actively studied by different
groups. An impetus has been given by the conference in Les Houches in
2002 \cite{dauxois} which showed the connections and the analogies
between different topics: astrophysics, two-dimensional hydrodynamics,
plasma physics,... Systems with long-range interactions have very
peculiar properties.  They can be spatially inhomogeneous (due to the
spontaneous formation of coherent structures), the ordinary
thermodynamic limit $N\rightarrow +\infty$ with fixed $N/V$ does not
apply and the statistical ensembles are generically inequivalent. This
does not mean that statistical mechanics breaks down for these
systems, but simply that it must be reformulated so as to take into
account their peculiarities. Therefore, we must go back to the
foundations and to the basic principles of statistical mechanics,
thermodynamics and kinetic theory.

In previous papers of this series \cite{hb1,hb2,hb3}, we have
developed a statistical mechanics and a kinetic theory adapted to
systems with weak long-range interactions. Our approach heavily relies
on many important works that have been developed in astrophysics for
stellar systems
\cite{saslaw,paddy,review}, in hydrodynamics for two-dimensional vortices
\cite{sommeria,tabeling,houches} and in 
plasma physics \cite{balescubook,pitaevskii}. However, the originality of our
approach is to remain as general as possible and develop a formalism
that can be applied to a wide variety of systems with long-range
interactions. This includes important toy models like the HMF model
\cite{hmf}, for example, where explicit analytical results can be
obtained. However, our approach is more general and aims at showing
the unity of the subject and the connection between different systems,
emphasizing their analogies and differences. In this sense, we  extend
our original approach
\cite{thesis,houches} where we first showed the analogy between the
statistical mechanics of stellar systems and two-dimensional vortices,
which are two physical systems of considerable interest.

In this paper, we consider material particles having inertia
\footnote{The case of 2D point vortices, that have no inertia, is special
and must be treated separately \cite{vortex}.} and interacting via a
{\it weak} long-range binary potential of interaction $u(|{\bf r}-{\bf
r}'|)$ in a space of dimension $d$. In Paper I, we have determined the
statistical equilibrium states and the static correlation functions in
a properly defined thermodynamic limit. For attractive potentials, we
have shown the existence of a critical energy $E_{c}$ (in the
microcanonical ensemble) or a critical temperature $T_{c}$ (in the
canonical ensemble) separating a spatially homogeneous phase from a
spatially inhomogeneous phase.  In Paper II, using an analogy with
plasma physics, we have developed a kinetic theory of systems with
long-range interactions in the spatially homogeneous phase. In Paper
III, we have studied the growth of correlations from the BBGKY
hierachy and the connection to the kinetic theory. In the present
paper, we further develop the kinetic theory of Hamiltonian systems
with long-range interactions by starting from the Klimontovitch
equation and using a quasilinear theory. We derive general kinetic
equations that can be applied to spatially inhomogeneous systems and
that take into account memory effects. These peculiarities are
specific to systems with unshielded long-range interactions and are
{\it novel} with respect to the much more studied case of spatially
homogeneous systems with short-range (or shielded)
interactions. However, we show that when the system is spatially
homogeneous and when memory effects can be neglected, we recover the
familiar kinetic equations of plasma physics
\footnote{The kinetic equations discussed in \cite{hb2} have the same
form as the Landau and Lenard-Balescu equations of plasma physics
except that the Fourier transform of the potential of interaction
$\hat{u}_{plasma}(k)\sim e^{2}/k^{2}$ is replaced by a more general
potential $\hat{u}(k)$ that can take negative values in the case of
{\it attractive} interactions. This change of sign is crucial for the
stability of the homogeneous phase and is responsible for
instabilities (similar to the Jeans instability) for $T<T_{c}$ or
$E<E_{c}$. In the case where the homogeneous phase is stable (for
$T>T_{c}$ or $E>E_{c}$), the potential $\hat{u}(k)$ enters explicitly
in the Lenard-Balescu equation (II-49) through the dielectric function
and the results can be different from those obtained with the
Coulombian potential $\hat{u}_{plasma}(k)$.  However, when collective
effects are ignored, we get the Landau equation (II-40) where the
potential of interaction $\hat{u}(k)$ appears only in a multiplicative
constant (II-43) controlling the timescale of the relaxation. Note finally
that the results of the kinetic theory  depend on the dimension of
space $d$
\cite{landaud}.} discussed in Paper II.

This paper is organized as follows. In Sec. \ref{sec_c}, we derive a
general kinetic equation from the Klimontovich equation by using a
quasilinear theory. This equation is valid at order $O(1/N)$ in the
proper thermodynamic limit $N\rightarrow +\infty$ defined in Paper I.
It coincides with the kinetic equation obtained from the BBGKY
hierarchy in Paper III.  For $N\rightarrow +\infty$, this kinetic
equation reduces to the Vlasov equation. At order $O(1/N)$ it takes
into account the effect of ``collisions'' (more properly
``correlations'') between particles due to finite $N$ effects. It
describes therefore the evolution of the system on a timescale $Nt_D$,
where $t_D$ is the dynamical time. This general kinetic equation
applies to systems that can be spatially inhomogeneous and takes into
account non-markovian effects. However, in order to obtain a closed
kinetic equation, we have been obliged to neglect some collective
effects. This is the main drawback of our approach: a more general
treatment should take into account both spatial inhomogeneity and
collective effects. If we restrict ourselves to spatially homogeneous
systems and neglect memory terms, we recover the Landau equation as a
special case. Therefore, the collective effects that we have neglected
correspond to the effects of polarization taken into account in the
Lenard-Balescu equation when the system is homogeneous (in plasma
physics, they lead to Debye shielding).  In Sec. \ref{sec_v}, we
develop a quasilinear theory of the Vlasov equation \cite{kp,sl,dub}
in relation with the process of violent relaxation \cite{lb,super} in
the collisionless regime of the dynamics. We derive a kinetic equation
for the coarse-grained distribution function $\overline{f}({\bf
r},{\bf v},t)$ and use this equation to describe the problem of {\it
incomplete relaxation}
\cite{next05} leading to deviations from the Lynden-Bell
distribution. We show the analogies and the differences between the
quasilinear theory of the Vlasov equation used to describe the process
of violent collisionless relaxation and the quasilinear theory of the
Klimontovich equation used to describe the process of slow collisional
relaxation.  In Sec. \ref{sec_test}, we consider the relaxation of a
test particle in a bath of field particles. The relaxation of the test
particle is due to the combined effect of a diffusion term and a
friction term. We derive the diffusion coefficient from the Kubo
formula and the friction term from a linear response theory based on
the Klimontovich equation. Like in the previous sections, the
originality of our approach is to develop a formalism that can
describe spatially inhomogeneous systems and that can take into
account memory terms.  If we consider spatially homogeneous systems
with short memory, we recover the results obtained in Paper
II. However, spatial inhomogeneity and memory effects can be important
in systems with long-range interactions. Therefore, in
Sec. \ref{sec_nm} we derive non-markovian kinetic equations that
generalize the standard Fokker-Planck equations.  We consider explicit
applications to self-gravitating systems and to the HMF model.

\section{Kinetic equations from a quasilinear theory}\label{sec_quasi}

In this section, we obtain a general kinetic equation (\ref{c13})
describing the collisional evolution of a Hamiltonian system of
particles with weak long-range interactions.  This equation, derived
from a quasilinear theory of the Klimontovich equation, is valid at
order $O(1/N)$ in the proper thermodynamics limit $N\rightarrow
+\infty$ defined in Paper I.  Then, we discuss the analogies and
the differences with the quasilinear theory of the Vlasov equation
developed in \cite{kp,sl,dub} to describe the process of violent
relaxation \cite{lb,super} in the collisionless regime.

\subsection{The slow collisional relaxation}\label{sec_c}

The exact distribution function (DF) of a system of particles in
interaction is a sum of Dirac functions
\begin{eqnarray}
f_{d}({\bf r},{\bf v},t)=\sum_{i}m\delta({\bf r}-{\bf
r}_{i}(t))\delta({\bf v}-{\bf v}_{i}(t)), \label{c1}
\end{eqnarray}
satisfying  the Klimontovich equation
\begin{eqnarray}
\frac{\partial f_{d}}{\partial t}+{\bf v}\frac{\partial
f_{d}}{\partial {\bf r}}-\nabla\Phi_d\frac{\partial f_{d}}{\partial
{\bf v}}=0, \label{c2}
\end{eqnarray}
where $\Phi_d({\bf r},t)=\int u(|{\bf r}-{\bf r}'|)f_{d}({\bf r}',{\bf
v}',t)d{\bf r}'d{\bf v}'$ is the exact potential created by $f_d$.
The Klimontovich equation (\ref{c2}) should not be confused with the
Vlasov equation (\ref{v1}) which has the same mathematical structure
but which applies to the {\it smooth} distribution function $f$. The
Vlasov equation is valid during the collisionless regime (see
Sec. \ref{sec_v}) while the Klimontovich equation is exact and
contains the same information as the Hamiltonian equations (I-1). We
now decompose the exact distribution function in the form
$f_{d}=f+\delta f$ where $f=\langle f_d\rangle$ is the smooth
distribution function and $\delta f$ the fluctuation around it.
Substituting this decomposition in Eq. (\ref{c2}) and locally
averaging over the fluctuations, we get
\begin{eqnarray}
\frac{\partial f}{\partial t}+Lf=\left\langle \nabla\delta\Phi\frac{\partial \delta f}{\partial {\bf v}}\right\rangle,
\label{c3}
\end{eqnarray}
where $L={\bf v}\frac{\partial}{\partial {\bf
r}}-\nabla\Phi\frac{\partial}{\partial {\bf v}}$ is an advection
operator in phase space constructed with the smooth field.
Subtracting Eq. (\ref{c3}) from Eq. (\ref{c2}) and neglecting non
linear terms in the fluctuations \footnote{As shown in Papers I and
III, the proper thermodynamic limit corresponds to $N\rightarrow
+\infty$ in such a way that the coupling constant $u_{*}\sim 1/N$
while the individual mass $m\sim 1$, the temperature $\beta\sim 1$,
the energy per particle $E/N\sim 1$ and the volume $V\sim 1$ are
fixed. This implies that $|{\bf r}|\sim 1$, $|{\bf v}|\sim 1$. We also
have $f/N\sim 1$ and $\delta f/N\sim 1/\sqrt{N}$ so that
$\Phi\sim u_{*}f\sim 1$ and $\delta\Phi\sim u_{*}\delta f\sim
1/\sqrt{N}$. With these scalings, we see that the terms that we have
kept in Eq. (\ref{c4}) are of order $\delta f\sim \sqrt{N}$ and
$f\delta\Phi \sim
\sqrt{N}$ while the nonlinear terms that we have neglected are of
order $\delta f\delta\Phi\sim 1\ll \sqrt{N}$. We also note that the
l.h.s. of Eq. (\ref{c3}) is of order $f\sim N$ while the r.h.s. of
Eq. (\ref{c3}) is of order $\delta f\delta\Phi \sim 1$. It would have
been more relevant to work in terms of the normalized distribution
function $F=f/N$. Then Eq. (\ref{c3}) can be rewritten
$\partial_{t}{F}+L{F}=(1/N)C(F)$ where the advective term is of order
$O(1)$ and the collision term is of order $1/N$. Therefore, this
equation describes the evolution of the system on a timescale $\sim
Nt_{D}$. For $N\rightarrow +\infty$, it reduces to the Vlasov
equation $\partial_{t}{F}+L{F}=0$.  }, we obtain the following equation
for the evolution of the fluctuations
\begin{eqnarray}
\frac{\partial \delta f}{\partial t}+L\delta f= \nabla\delta\Phi\frac{\partial f}{\partial {\bf v}}.
\label{c4}
\end{eqnarray}
Equations (\ref{c3}) and (\ref{c4}) form the basis of the quasilinear
theory. For spatially homogeneous systems, they can be solved with the
aid of Laplace-Fourier transforms and they yield the Lenard-Balescu
equation (see, e.g., \cite{pitaevskii} and Appendix B of Paper II).  In the
present work, we shall proceed differently so as to treat the case of
systems that are not necessarily spatially homogeneous and not
necessarily markovian. Our method avoids the use of Laplace-Fourier
transforms and remains in physical space. This yields expressions with
a clear interpretation which enlightens the basic physics. The
drawback of our approach, however, is that it neglects collective
effects. The formal solution of Eq. (\ref{c4}) is
\begin{eqnarray}
\delta f(t)=G(t,0)\delta f(0)+\int_{0}^{t}d\tau
G(t,t-\tau)\nabla\delta\Phi (t-\tau)\frac{\partial f}{\partial {\bf
v}}(t-\tau), \label{c5}
\end{eqnarray}
where $G$ is the Green function associated with the advection
operator $L$ and we have noted $f(t)=f({\bf r},{\bf v},t)$ and
$\delta\Phi(t)=\delta\Phi({\bf r},t)$ for brevity.  On the other
hand, the perturbation of the potential is related to the
perturbation of the distribution function through
\begin{eqnarray}
-\nabla\delta\Phi(t)=\frac{1}{m}\int {\bf F}(1\rightarrow 0)\delta
f_{1}(t)d{\bf x}_{1}, \label{c6}
\end{eqnarray}
where $0$ refers to the position ${\bf r}$ and we have noted $\delta
f_{1}(t)=\delta f({\bf r}_{1},{\bf v}_{1},t)$. Therefore, considering
Eqs. (\ref{c5}) and (\ref{c6}), we see that the fluctuation of the
field $\nabla\delta\Phi(t)$ is given by an iterative process:
$\nabla\delta\Phi(t)$ depends on $\delta f_{1}(t)$ which itself
depends on $\nabla\delta\Phi_{1}(t-\tau)$ etc. We shall solve this
problem perturbatively in the thermodynamic limit $N\rightarrow
+\infty$. To leading order, we get
\begin{eqnarray}
\left\langle \nabla\delta\Phi\frac{\partial \delta f}{\partial {\bf v}}\right\rangle=-\frac{1}{m}\frac{\partial}{\partial v^{\mu}}\int d{\bf x}_{1}F^{\mu}(1\rightarrow 0)G_{1}(t,0)G(t,0)\langle \delta f_{1}(0)\delta f(0)\rangle\nonumber\\
+\frac{1}{m^{2}}\frac{\partial}{\partial v^{\mu}}\int_{0}^{t}d\tau\int d{\bf x}_{1}d{\bf x}_{2}F^{\mu}(1\rightarrow 0)
G_{1}(t,t-\tau)G(t,t-\tau)\nonumber\\
\times\biggl\lbrace F^{\nu}(2\rightarrow 0)\langle \delta f_{1}(t-\tau)\delta f_{2}(t-\tau)\rangle \frac{\partial f}{\partial v^{\nu}}(t-\tau)\nonumber\\
+ F^{\nu}(2\rightarrow 1)\langle \delta f(t-\tau)\delta f_{2}(t-\tau)\rangle \frac{\partial f_{1}}{\partial v_{1}^{\nu}}(t-\tau)\biggr\rbrace.\nonumber\\
\label{c7}
\end{eqnarray}
 Now, the fluctuation is exactly defined by
\begin{eqnarray}
\delta f({\bf r},{\bf v},t)=\sum_{i}m\delta({\bf r}-{\bf
r}_{i}(t))\delta({\bf v}-{\bf v}_{i}(t))-f({\bf r},{\bf v},t).
\label{c8}
\end{eqnarray}
Therefore, we obtain
\begin{eqnarray}
\langle \delta f_{1}\delta f_{2}\rangle =\langle \sum_{i\neq j}m^{2}\delta({\bf x}_{1}-{\bf x}_{i})\delta({\bf x}_{2}-{\bf x}_{j})\rangle +\langle \sum_{i}m^{2}\delta({\bf x}_{1}-{\bf x}_{i})\delta({\bf x}_{2}-{\bf x}_{i})\rangle \nonumber\\
-\langle \sum_{i}m\delta({\bf x}_{1}-{\bf x}_{i})f_{2}\rangle  -\langle  \sum_{j}m\delta({\bf x}_{2}-{\bf x}_{j})f_{1}\rangle+f_{1}f_{2}.
\label{c9}
\end{eqnarray}
To evaluate the correlation function, we average with respect to the smooth
distribution $f_{i}/(Nm)$ or $f_{i}f_{j}/(Nm)^{2}$.
This operation leads to
\begin{eqnarray}
\langle \delta f_{1}\delta f_{2}\rangle =\frac{N-1}{N}f_{1}f_{2} +mf_{1}\delta({\bf x}_{1}-{\bf x}_{2})
-f_{1}f_{2}-f_{2}f_{1}+f_{1}f_{2},
\label{c10}
\end{eqnarray}
so that, finally,
\begin{eqnarray}
\langle \delta f_{1}\delta f_{2}\rangle =mf_{1}\delta({\bf x}_{1}-{\bf x}_{2})
-\frac{1}{N}f_{1}f_{2}.
\label{c11}
\end{eqnarray}
Substituting this result in Eq. (\ref{c7}), we find that
\begin{eqnarray}
\left\langle \nabla\delta\Phi\frac{\partial \delta f}{\partial {\bf v}}\right\rangle=\langle F^{\mu}(1\rightarrow 0)\rangle\frac{\partial f}{\partial v^{\mu}}
+\frac{1}{m}\frac{\partial}{\partial v^{\mu}}\int_{0}^{t}d\tau\int d{\bf x}_{1}{F}^{\mu}(1\rightarrow 0)
G(t,t-\tau)\nonumber\\
\times\biggl\lbrace {\cal F}^{\nu}(1\rightarrow 0)f_{1}(t-\tau)\frac{\partial f}{\partial v^{\nu}}(t-\tau)
+ {\cal F}^{\nu}(0\rightarrow 1)f(t-\tau)\frac{\partial f_{1}}{\partial v_{1}^{\nu}}(t-\tau)\biggr\rbrace,
\label{c12}
\end{eqnarray}
where we have regrouped the two Greenians $G$ and $G_1$ in a single
notation for brevity. Finally, replacing this expression in Eq. (\ref{c3}), we
obtain the kinetic equation
\begin{eqnarray}
\frac{\partial f}{\partial t}+{\bf v}\frac{\partial f}{\partial {\bf r}}+\frac{N-1}{N}\langle {\bf F}\rangle \frac{\partial f}{\partial {\bf v}}=m \frac{\partial}{\partial v^{\mu}}\int_{0}^{t}d\tau\int d{\bf r}_{1}d{\bf v}_{1}\frac{{F}^{\mu}}{m}(1\rightarrow 0)
G(t,t-\tau)\nonumber\\
\times \biggl\lbrace \frac{{\cal F}^{\nu}}{m}(1\rightarrow 0)f_1\frac{\partial f}{\partial v^{\nu}}
+ \frac{{\cal F}^{\nu}}{m}(0\rightarrow 1)f\frac{\partial f_1}{\partial v_{1}^{\nu}}\biggr\rbrace_{t-\tau}.
\label{c13}
\end{eqnarray}
This is identical to the general kinetic equation (33) of Paper III
obtained from the BBGKY hierarchy (or from the projection operator
formalism \cite{kandrup1}).  We note that the term of order $1/N$ in
the l.h.s. comes from the first term in Eq. (\ref{c5}). It corresponds
to the mere advection of the fluctuations by the smooth field in
Eq. (\ref{c4}), i.e. ignoring the coupling between the fluctuations of
the field and the smooth distribution function (r.h.s. of
Eq. (\ref{c4})) which gives rise to the collision term.

\subsection{The violent collisionless relaxation}\label{sec_v}

To leading order in $N\rightarrow +\infty$, the smooth distribution
function $f({\bf r},{\bf v},t)$ is solution of the Vlasov equation
\begin{eqnarray}
\frac{\partial f}{\partial t}+{\bf v}\frac{\partial f}{\partial {\bf
r}}-\nabla\Phi\frac{\partial f}{\partial {\bf v}}=0, \label{v1}
\end{eqnarray}
where $\Phi({\bf r},t)=\int u(|{\bf r}-{\bf r}'|)f({\bf r}',{\bf
v}',t)d{\bf r}'d{\bf v}'$ is the smooth potential created by $f$.  The
Vlasov equation describes the collisionless evolution of the system
due to mean field effects only, before the cumulative nature of
the collisions becomes manifest on a timescale $t_{coll}\sim Nt_D$ or
larger. Starting from an initial condition which is dynamically
unstable, the Vlasov equation coupled to a long-range potential of
interaction develops an intricate filamentation in phase space at
smaller and smaller scales. In this sense, the fine-grained
distribution function $f({\bf r},{\bf v},t)$ never achieves
equilibrium. However, if we locally average over the filaments, the
resulting ``coarse-grained'' distribution function $\overline{f}({\bf
r},{\bf v},t)$ will achieve a steady state on a timescale $\sim
t_{D}$. Since the Vlasov equation is only valid in the collisionless
regime $t\ll t_{coll}$, this corresponds to a quasi-stationary state
(QSS) that will slowly evolve under the effect of collisions on a 
timescale $\sim Nt_{D}$ or larger. We can try to predict this QSS in
terms of a statistical mechanics of the Vlasov equation, using the
approach of Lynden-Bell
\cite{lb} developed for collisionless stellar systems (see also
\cite{super}).  In the case where the fine-grained distribution
function $f({\bf r},{\bf v},t)$ takes only two values $0$ and
$\eta_{0}$, the statistical equilibrium state maximizes the
Lynden-Bell entropy
\begin{eqnarray}
S_{L.B.}=-\int \biggl\lbrace \frac{\overline{f}}{\eta_{0}}\ln \frac{\overline{f}}{\eta_{0}}+\biggl (1- \frac{\overline{f}}{\eta_{0}}\biggr )\ln \biggl (1-\frac{\overline{f}}{\eta_{0}}\biggr )\biggr\rbrace d{\bf r}d{\bf v},
\label{v2}
\end{eqnarray}
at fixed mass and energy. This leads to the coarse-grained
distribution function
\begin{eqnarray}
\overline{f}=\frac{\eta_{0}}{1+ e^{\beta\eta_{0}(\frac{v^{2}}{2}+\Phi)-\mu}}. 
\label{v3}
\end{eqnarray}
Note that the mixing entropy (\ref{v2}) is formally similar to the
Fermi-Dirac entropy and the equilibrium distribution (\ref{v3}) is
formally similar to the Fermi-Dirac distribution. An effective
``exclusion principle'', similar to the Pauli principle in quantum
mechanics, arises in the theory of violent relaxation because the
different phase levels cannot overlap. We stress that the Lynden-Bell
theory is based on an assumption of ergodicity. Indeed, it implicity
assumes that the phase elements mix efficiently during the dynamics so that
the QSS is the {\it most mixed state} compatible with the integral
constraints of the Vlasov equation. This may not always be the case as
discussed in the sequel.

We can try to determine the dynamical equation satisfied by the
coarse-grained distribution function $\overline{f}({\bf r},{\bf
v},t)$ by developing a quasilinear theory of the Vlasov equation. We
decompose the distribution function in the form
$f=\overline{f}+\tilde f$ where $\overline{f}$ is the coarse-grained
distribution function and $\tilde f\ll \overline{f}$ a fluctuation around it.
Substituting this decomposition in Eq. (\ref{v1}) and taking the local
average, we get
\begin{equation}
\label{v4} {\partial \overline{f}\over\partial
t}+L\overline{f}={\partial \over \partial {\bf
v}}\nabla\overline{\tilde\Phi\tilde f},
\end{equation}
where $L={\bf v}\frac{\partial}{\partial {\bf
r}}-\nabla\Phi\frac{\partial}{\partial {\bf v}}$ is an advection
operator in phase space constructed with the smooth field.
Subtracting Eq. (\ref{v4}) from Eq. (\ref{v1}) and neglecting
nonlinear terms in the fluctuations, we obtain an equation for the
perturbation
\begin{equation}
\label{v5} {\partial \tilde{f}\over\partial
t}+L\tilde{f}=\nabla\tilde{\Phi}{\partial\overline{f}\over\partial
{\bf v}}.
\end{equation}
Equations (\ref{v4}) and (\ref{v5}) are formally similar to
Eqs. (\ref{c3}) and (\ref{c4}) of the previous section but with a
completely different interpretation.  In Sec. \ref{sec_c}, the subdynamics was
played by $f_{d}$ (a sum of $\delta$-functions) and the macrodynamics
by $f$ (a smooth field).  The smooth field averages over the positions
of the $\delta$-functions that strongly fluctuate. In the phase of
violent relaxation, the ``smooth'' field $f$ develops itself a finely
striated structure and strongly fluctuates. Therefore, it is {\it not}
smooth at a higher scale of resolution and a second smoothing
procedure (coarse-graining) must be introduced. In that case, the
subdynamics is played by $f$ and the macrodynamics by
$\overline{f}$. The coarse-grained field averages over the positions
of the filaments.  

The coupled equations (\ref{v4}) and (\ref{v5}) can be solved by an
iterative procedure similar to that developed in Sec. \ref{sec_c} and
we finally obtain
\begin{eqnarray}
\label{v6} {\partial \overline{f}\over\partial
t}+L\overline{f}={\partial \over \partial
v^{\mu}}\int_{0}^{t} d\tau\int d{\bf r}_{1}
d{\bf v}_{1} d{\bf r}_{2}d{\bf v}_{2} \frac{F^{\mu}}{m}(1\rightarrow 0)G_{1}(t,t-\tau)G(t,t-\tau)\nonumber\\
\times\biggl\lbrace \frac{F^{\nu}}{m}(2\rightarrow 0)\overline{ \tilde f({\bf r}_{1},{\bf v}_{1},t-\tau)\tilde f({\bf r}_{2},{\bf v}_{2},t-\tau)} {\partial \overline{f}\over \partial v^{\nu}}({\bf r},{\bf v},t-\tau)\nonumber\\
+\frac{F^{\nu}}{m}(2\rightarrow 1)\overline{\tilde f({\bf r},{\bf
v},t-\tau)\tilde f({\bf r}_2,{\bf v}_2,t-\tau)} {\partial
\overline{f}\over \partial v_{1}^{\nu}}({\bf r}_{1},{\bf
v}_{1},t-\tau) \biggr \rbrace.
\end{eqnarray}
To close the system, it remains for one to evaluate the correlation
function $\overline{\tilde f({\bf r},{\bf v},t)\tilde f({\bf
r}_1,{\bf v}_1,t)}$. We shall assume that the mixing in phase space is
sufficiently efficient so that the scale of the kinematic
correlations is small with respect to the coarse-graining mesh size.
In that case,
\begin{equation}
\label{v7} \overline{\tilde f({\bf r},{\bf v},t)\tilde f({\bf
r}_{1},{\bf v}_{1},t)}=\epsilon_{r}^{d}\epsilon_{v}^{d} \delta({\bf
r}-{\bf r}_{1})\delta ({\bf v}-{\bf v}_{1}) \overline{\tilde f^{2}}({\bf
r},{\bf v},t),
\end{equation}
where $\epsilon_{r}$ and $\epsilon_{v}$ are the resolution scales in
position and velocity respectively. Now,
\begin{equation}
\label{v8}
 \overline{\tilde f^{2}}=\overline{(f-\overline{f})^{2}}=\overline{f^{2}}-\overline{f}^{2}.
\end{equation}
We shall assume, for simplicity, that the initial condition in phase
space consists of patches where the distribution function takes a
unique value $f=\eta_{0}$ surrounded by vacuum ($f=0$). In this
two-levels approximation $\overline{f^{2}}=\overline{\eta_{0}\times
f}=
\eta_{0}\overline{f}$ and, therefore,
\begin{equation}
\label{v9} \overline{\tilde f({\bf r},{\bf v},t)\tilde f({\bf
r}_{1},{\bf v}_{1},t)}=\epsilon_{r}^{d}\epsilon_{v}^{d} \delta({\bf
r}-{\bf r}_{1})\delta ({\bf v}-{\bf v}_{1}) \overline{f}
(\eta_{0}-\overline{f}).
\end{equation}
Substituting this expression in Eq. (\ref{v6}) and carrying
out the integrations on ${\bf r}_2$ and ${\bf v}_2$, we obtain
\begin{eqnarray}
\label{v10} {\partial \overline{f}\over\partial
t}+L\overline{f}={\epsilon_{r}^{d}\epsilon_{v}^{d}}
{\partial \over
\partial v^{\mu}}\int_{0}^{t} d\tau\int d{\bf r}_{1}d{\bf v}_{1}
\frac{F^{\mu}}{m}(1\rightarrow 0) G(t,t-\tau) \frac{F^{\nu}}{m}(1\rightarrow 0)\nonumber\\
\times\biggl\lbrace  \overline{f}_1(\eta_{0}-\overline{f}_1){\partial
\overline{f}\over \partial v^{\nu}} -
\overline{f}(\eta_{0}-\overline{f}) {\partial \overline{f}_1\over
\partial v_1^{\nu}} \biggr \rbrace_{t-\tau},
\end{eqnarray}
where $f=f({\bf r},{\bf v},t-\tau)$ and $f_1=f({\bf r}_1,{\bf
v}_1,t-\tau)$. This equation is expected to describe the late
quiescent stages of the relaxation process when the fluctuations have
weaken so that the quasilinear approximation can be implemented. It
does not describe the early very chaotic process of violent relaxation
driven by the strong fluctuations of the potential. The quasilinear
theory of the Vlasov equation is therefore a theory of ``quiescent''
collisionless relaxation.

Equation (\ref{v10}) is similar, in structure, to Eq. (\ref{c13})
describing the collisional evolution of the system with, nevertheless,
three important differences: (i) the fluctuating force ${\cal
F}(1\rightarrow 0)$ is replaced by the direct force ${F}(1\rightarrow 0)$
because the fluctuations are taken into account differently.  (ii) The
distribution function $f$ in the collisional term of Eq.  (\ref{c13})
is replaced by the product $\overline{f}(\eta_0-\overline{f})$ in
Eq. (\ref{v10}). This nonlinear term arises from the effective
``exclusion principle'', discovered by Lynden-Bell, accounting for the
non-overlapping of phase levels in the collisionless regime. This is
consistent with the Fermi-Dirac-like entropy (\ref{v2}) and
Fermi-Dirac-like distribution (\ref{v3}) at statistical equilibrium
(iii) Considering the dilute limit $\overline{f}\ll\eta_{0}$ to fix
the ideas, we see that the equations (\ref{v10}) and (\ref{c13}) have
the same mathematical form differing only in the prefactors: the mass
$m$ of a particle in Eq. (\ref{c13}) is replaced by the mass
$\eta_{0}\epsilon_r^{d}\epsilon_v^d$ of a completely filled macrocell
in Eq. (\ref{v10}). This implies that the timescales of collisional
and collisionless relaxation are in the ratio
\begin{eqnarray}
\frac{t_{ncoll}}{t_{coll}}\sim
\frac{m}{\eta_{0}\epsilon_r^{d}\epsilon_v^d}. \label{v12}
\end{eqnarray}
Since $\eta_{0}\epsilon_{r}^{d}\epsilon_{v}^{d}\gg m$, this ratio is
in general quite small implying that the collisionless relaxation is
much more rapid than the collisional relaxation. Typically,
$t_{ncoll}$ is of the order of a few dynamical times $t_{D}$ (its
precise value depends on the size of the mesh) while $t_{coll}$ is of
order $\sim {N}t_{D}$ or larger. The kinetic equation (\ref{v10})
conserves the mass and, presumably, the energy. By contrast, we cannot
prove an $H$-theorem for the Lynden-Bell entropy (\ref{v2}). Indeed,
the time variation of the Lynden-Bell entropy is of the form
\begin{eqnarray}
\dot S_{L.B.}=\frac{1}{2}\epsilon_{r}^{d}\epsilon_{v}^{d}\int d{\bf r}d{\bf v}d{\bf r}_{1}d{\bf v}_{1}\frac{1}{\overline{f}(\eta_{0}-\overline{f})\overline{f}_{1}(\eta_{0}-\overline{f}_{1})}\int_{0}^{t}d\tau Q(t) G(t,t-\tau)Q(t-\tau),\label{g6fgh}
\end{eqnarray}
\begin{eqnarray}
Q(t)=\frac{{F}^{\mu}}{m}(1\rightarrow 0,t) \left\lbrack \overline{f}_{1}(\eta_{0}-\overline{f}_{1})\frac{\partial \overline{f}}{\partial v^{\mu}}-\overline{f}(\eta_{0}-\overline{f}) \frac{\partial \overline{f}_{1}}{\partial v_{1}^{\mu}}\right\rbrack,\label{g6fghb}
\end{eqnarray}
and its sign is not necessarily positive.  This depends on the
importance of memory terms. In addition, even if Eq. (\ref{v10})
conserves energy and increases the Fermi-Dirac entropy monotonically,
this does not necessarily imply that the system will converge towards
the Lynden-Bell distribution (\ref{v3}). It has been observed in
several experiments and numerical simulations that the QSS does not
coincide with the statistical equilibrium state predicted by
Lynden-Bell. This {\it incomplete relaxation} \cite{next05} is usually
explained by a lack of ergodicity and ``incomplete mixing''. In fact,
very few is known concerning kinetic equations of the form of
Eq. (\ref{v10}) and it is not clear whether the Lynden-Bell
distribution (\ref{v3}) is a stationary solution of that equation (and
if it is the only one). As explained in Paper III for the kinetic
equation (\ref{c13}) describing the collisional relaxation, the
relaxation may stop because the current ${\bf J}$ vanishes due to the
{\it absence of resonances}. This argument may also apply to
Eq. (\ref{v10}) which has a similar structure and can be a cause for
incomplete relaxation. The system tries to approach the statistical
equilibrium state (as indicated by the increase of the entropy) but
may be trapped in a QSS that is different from the statistical
prediction (\ref{v3}). This QSS is a steady solution of Eq.
(\ref{v10}), or more generally (\ref{v6}), which cancels individually
the advective term (l.h.s.) and the effective collision term (r.h.s.).
This determines a subclass of steady states of the Vlasov equation
(cancellation of the l.h.s.)  such that the complicated ``turbulent''
current ${\bf J}$ in the r.h.s. vanishes.  This offers a large class
of possible steady state solutions that can explain the deviation
between the QSS and the Lynden-Bell statistical equilibrium state
(\ref{v3}) observed, in certain cases, in simulations and experiments
of violent relaxation. Other causes of incomplete relaxation, due to
the rapid decay of the fluctuations in space and time (leading to a
small value of the current), will be described in Sec. \ref{sec_case}.

\subsection{The case of stellar systems}\label{sec_case}

The case of stellar systems is special and deserves a specific
discussion. These systems are spatially inhomogeneous but, due to the
divergence of the gravitational force ${\bf F}(1\rightarrow 0)$ when
${\bf r}_{1}\rightarrow {\bf r}$, the integral in the r.h.s. of
Eq. (\ref{v10}) can be evaluated by making a {\it local approximation}
which amounts to replacing $f({\bf r}_1,{\bf v}_1,t)$ by $f({\bf
r},{\bf v}_1,t)$. This approximation is justified by the fact that the
diffusion coefficient diverges logarithmically when ${\bf
r}_{1}\rightarrow {\bf r}$ (see below).  We shall also make a
markovian approximation $f({\bf r},{\bf v}_1,t-\tau)\simeq f({\bf
r},{\bf v}_1,t)$, $f({\bf r},{\bf v},t-\tau)\simeq f({\bf r},{\bf
v},t)$ and extend the time integration to $+\infty$. Then,
Eq. (\ref{v10}) becomes
\begin{eqnarray}
\label{v10b} {\partial \overline{f}\over\partial
t}+L\overline{f}={\epsilon_{r}^{3}\epsilon_{v}^{3}}
{\partial \over
\partial v^{\mu}}\int_{0}^{+\infty} d\tau\int d{\bf r}_{1}d{\bf v}_{1}
\frac{F^{\mu}}{m}(1\rightarrow 0,t)  \frac{F^{\nu}}{m}(1\rightarrow 0,t-\tau)\nonumber\\
\times\biggl\lbrace  \overline{f}_1(\eta_{0}-\overline{f}_1){\partial
\overline{f}\over \partial v^{\nu}} -
\overline{f}(\eta_{0}-\overline{f}) {\partial \overline{f}_1\over
\partial v_1^{\nu}} \biggr \rbrace,
\end{eqnarray}
where now $f=f({\bf r},{\bf v},t)$ and $f_1=f({\bf r},{\bf
v}_1,t)$. Making a linear trajectory approximation ${\bf
v}_{i}(t-\tau)={\bf v}_{i}(t)$ and ${\bf r}_{i}(t-\tau)={\bf
r}_{i}-{\bf v}_{i}\tau$, we can perform the integrations on ${\bf
r}_{1}$ and $\tau$ like in Appendix A of Paper II. This yields the generalized
Landau equation
\begin{eqnarray}
\frac{\partial \overline{f}}{\partial t}+L\overline{f}=\pi (2\pi)^3
\epsilon_{r}^{3}\epsilon_{v}^{3} \frac{\partial}{\partial {v}^{\mu}}
\int d{\bf v}_{1} d{\bf k} k^{\mu}k^{\nu} \hat{u}(k)^2 \delta({\bf
k}\cdot {\bf w}) \biggl\lbrace
\overline{f}_{1}(\eta_{0}-\overline{f}_{1}){\partial\overline{f}\over\partial
v^{\nu}}-\overline{f}(\eta_{0}-\overline{f}){\partial\overline{f}_{1}\over\partial
v_{1}^{\nu}}\biggr\rbrace.\nonumber\\
\label{v11}
\end{eqnarray}
As a result of the local approximation, the effect of the spatial
inhomogeneity is only retained in the advective term $L$ in the l.h.s.
of Eq. (\ref{v11}). The same approximations are made for collisional
stellar systems leading to Eq. (II-44) of Paper II. Equation
(\ref{v11}) can also be written in the form
\begin{eqnarray}
\frac{\partial \overline{f}}{\partial t}+L\overline{f}=
2\pi G^{2}\epsilon_{r}^{3}\epsilon_{v}^{3}\ln\Lambda\frac{\partial}{\partial {v}^{\mu}}
\int d{\bf v}_{1} \frac{w^{2}\delta^{\mu\nu}-w^{\mu}w^{\nu}}{w^{3}} \biggl\lbrace
\overline{f}_{1}(\eta_{0}-\overline{f}_{1}){\partial\overline{f}\over\partial
v^{\nu}}-\overline{f}(\eta_{0}-\overline{f}){\partial\overline{f}_{1}\over\partial
v_{1}^{\nu}}\biggr\rbrace,
\label{v11b}
\end{eqnarray}
where $\ln\Lambda=\int_{0}^{+\infty}dk/k$ is the Coulombian factor. It
exibits a logarithmic divergence at small and large scales and it must
be regularized by introducing some cut-offs, writing
$\ln\Lambda=\ln(L_{max}/L_{min})$. The integral at large scales must
be cut-off at the system size $L_{max}\sim R$ (or Jeans length) which
plays the role of the Debye length in the present context (see Paper
III). For collisional stellar systems, the integral at small scales
must be cut-off at the Landau length $L_{min}\sim Gm/v_{typ}^{2}$ with
$v_{typ}^{2}\sim GM/R$ corresponding to a deflexion at $90^{o}$ of the
particles' trajectory. This yields a
Coulomb factor $\ln\Lambda\sim \ln N$. In the present
context, the integral at small scales must be cut-off at the
resolution length $\epsilon_{r}$. Therefore,
$\ln\Lambda=\ln(R/\epsilon_{r})$. This implies that the timescale of
collisional relaxation and the timescale of violent relaxation are in
the ratio
\begin{eqnarray}
\frac{t_{ncoll}}{t_{coll}}\sim
\frac{m}{\eta_{0}\epsilon_r^{3}\epsilon_v^3}\frac{\ln(R/\epsilon_{r})}{\ln N}. 
\label{v11bede}
\end{eqnarray}

It is easy to check \cite{landaugen} that Eq.  (\ref{v11b}) conserves
the mass and the energy, that it monotonically increases the
Lynden-Bell entropy (\ref{v2}) ($H$-theorem) and that its only
stationary solution is the Lynden-Bell distribution
(\ref{v3}). Therefore, the kinetic equation (\ref{v11b}) tends to
reach the Lynden-Bell distribution (\ref{v3}). However, there are
several reasons why it cannot attain it: (i) {\it Evaporation:} for
self-gravitating systems, it is well-known that the Lynden-Bell
distribution (\ref{v3}) coupled to the Poisson equation has infinite
mass so that there is no physical distribution of the form (\ref{v3})
in an infinite domain.  The system can increase the Lynden-Bell
entropy indefinitely by evaporating. Therefore, the generalized
Vlasov-Landau equation (\ref{v11b}) has no steady state with finite
mass and the distribution function tends to spreads indefinitely.
(ii) {\it Incomplete relaxation in space:} The turbulent current ${\bf
J}$ in Eq. (\ref{v11b}), or more generally in Eq. (\ref{v6}), is
driven by the fluctuations $f_{2}\equiv \overline{\tilde{f}^{2}}$ of
the distribution function (generating the fluctuations $\delta\Phi$ of
the potential). In the ``mixing region'' of phase space where the
fluctuations are strong, the DF tends to reach the Lynden-Bell
distribution (\ref{v3}). As we depart from the ``mixing region'', the
fluctuations decay ($f_{2}\rightarrow 0$) and the mixing is less and
less efficient $\|{\bf J}\|\rightarrow 0$. In these regions, the
system takes a long time to reach the Lynden-Bell distribution
(\ref{v3}) and, in practice, cannot attain it in the time available
(see (iii)). In the two levels case, we have
$f_{2}=\overline{f}(\eta_{0}-\overline{f})$. Therefore, the phase
space regions where $\overline{f}\rightarrow 0$ or
$\overline{f}\rightarrow \eta_{0}$ do not mix well (the diffusion
current ${\bf J}$ is weak) and the observed DF can be sensibly
different from the Lynden-Bell distribution in these regions of phase
space. This concerns essentially the core ($\overline{f}\rightarrow
\eta_{0}$) and the tail ($\overline{f}\rightarrow 0$) of the
distribution. (iii) {\it Incomplete relaxation in time:} during
violent relaxation, the system tends to approach the statistical
equilibrium state (\ref{v3}). However, as it approaches equilibrium,
the fluctuations of the gravitational field, which are the engine of
the evolution, become less and less effective to drive the
relaxation. This is because the scale of the fluctuations becomes
smaller and smaller as time goes on. This effect can be taken into
account in the kinetic theory by considering that the correlation
lengths $\epsilon_{r}(t)$ and $\epsilon_{v}(t)$ decrease with time so
that, in the kinetic equation (\ref{v11b}), the prefactor
$\epsilon_{r}(t)\epsilon_{v}(t)\rightarrow 0$ for $t\rightarrow +\infty$.
As a result, the ``turbulent'' current ${\bf J}$ in Eq. (\ref{v11b})
can vanish {\it before} the system has reached the statistical equilibrium
state (\ref{v3}). In that case, the system can be trapped in a QSS that is a
steady solution of the Vlasov equation different from the statistical
prediction (\ref{v3}). Similar arguments have been given in \cite{csr}
on the basis of a more phenomenological kinetic theory of violent relaxation.
On longer timescale, the encounters must be taken into account. Then
the system is described by a collisional kinetic Vlasov-Landau
equation of the form (III-41). This equation conserves the mass, the
energy (kinetic $+$ potential) and monotonically increases the
Boltzmann entropy. The mean field Maxwell-Boltzmann distribution
(I-24) is the only stationary solution of this equation so that the
system tends to reach this distribution on a timescale $(N/\ln
N)t_{D}$. In practice, however, the convergence to the Boltzmann distribution
is hampered by the {\it escape of stars} and by the {\it gravothermal
catastrophe} \cite{saslaw,paddy,review}.

\subsection{Physical interpretation of the QSS}\label{sec_int}

Based on the preceding kinetic theory, we propose the following
interpretation \cite{houches} of the QSS observed in Hamiltonian systems with
long-range interactions:

1. The QSS results from a process of phase mixing and violent
relaxation.  This is a purely collisionless process driven by the
fluctuations of the mean-field potential. It takes place on a
timescale of a few dynamical times where the Vlasov equation is
valid. The QSS is a nonlinearly dynamically stable stationary solution
of the Vlasov equation on the coarse-grained scale, i.e. the {\it
coarse-grained} DF $\overline{f}_{QSS}({\bf r},{\bf v})$ is a stable
stationary solution of the Vlasov equation.  Since the Vlasov equation
admits an infinite number of stationary solutions, it is not easy to
predict the one which will be dynamically selected by the process of
violent relaxation.

2. In principle, the distribution $\overline{f}_{QSS}({\bf r},{\bf
v})$ of the QSS can be predicted from the statistical theory of the
Vlasov equation developed by Lynden-Bell \cite{lb}. The distribution
$\overline{f}_{L.B.}({\bf r},{\bf v})$ depends on the details of the
initial condition (in addition to the value of the mass and the
energy) because of the conservation of the Casimir constraints
\cite{super}. The coarse-grained DF predicted by Lynden-Bell looks
like a sort of superstatistics.

3. In many cases, the prediction of Lynden-Bell works well
\cite{hohl,rsmepp,staquet,antoniazzi1}. In certain cases, the prediction of
Lynden-Bell fails because of the complicated problem of {\it
incomplete relaxation}
\cite{next05}. The system tends to reach the Lynden-Bell distribution 
(as implied by the increase of the Lynden-Bell entropy) but cannot
attain it because the fluctuations of the potential (which drive the
evolution) fade away before the system has reached the most mixed
state. Therefore, the incompleteness of the violent relaxation is of
dynamical origin. In such cases, the QSS can take forms that are
different from the statistical prediction,
i.e. $\overline{f}_{QSS}\neq\overline{f}_{L.B.}$. Thus, other
distributions, that are stable stationary solutions of the Vlasov
equation, can emerge. For example, the Tsallis distributions
\cite{tsallis}  are {\it particular} stationary solutions
of the Vlasov equation (polytropes) \cite{cstsallis} that can
sometimes be reached as a result of an incomplete violent
relaxation. Several examples have been exhibited where the QSS
\cite{boghosian,brands,lrt,campa} or the transient stages  of the collisional relaxation \cite{ts,yamaguchi,campa} are remarkably
well fitted by Tsallis distributions (see the detailed discussion of
Paper III \cite{hb3}). This suggests that Tsallis distributions may
represent ``attractors'' of the Vlasov equation in case of incomplete
relaxation, for some particular initial conditions. However, they are
not ``universal attractors''
\footnote{Tsallis entropies apply when the phase space of a system is
fractal or multi-fractal. The fractal properties of the process of
violent relaxation are not known. For the HMF model, an interesting
regime where Tsallis thermodynamics seems to apply \cite{plr} has been
found above a critical magnetization \cite{tlb}.}. Indeed, other
distributions have been observed that differ both from the Lynden-Bell and
the Tsallis distributions. This is clear for galaxies in astrophysics
that are neither isothermal nor polytropic \cite{bt}. There are also
cases where the system does not reach a QSS and develops instead
long-lasting oscillations \cite{mineau,mk}. It would be interesting to
know whether these different possible behaviours are captured by the
kinetic equation (\ref{v10}).

4. Since the Lynden-Bell/Vlasov approach is restricted to the
Boltzmann $\mu$-space, that is only a projection of the full Gibbs
$\Gamma$-space, one could fear that some fundamental properties of the
latter (fractal structures, etc...) could be lost in that approach. In
fact, we believe that the Vlasov equation correctly describes the
regime where the QSS appears. Therefore, in this regime, all the
physics of the problem is contained in the Vlasov equation evolving in
$\mu$-space. However, the Vlasov equations is a very complicated
equation (like the Euler equations of turbulence for example). In
particular, it can exhibit fractal structures and non-ergodic
behaviours just as the $N$-body system does. Therefore, the Vlasov
equation is not in contradiction with a complex structure of phase
space: the striking features that have been observed for
the $N$-body problem such as QSS
\cite{lrt,yamaguchi,antoniazzi1,campa}, phase-space holes/clumps
\cite{mineau,mk}, anomalous diffusion
\cite{plr}, non-ergodic 
behaviours etc. should also be observed with the Vlasov equation
(except if they are due to finite $N$-effects which is also a
possibility to consider).

\section{Relaxation of a test particle in a bath}\label{sec_test}

In this section, we study the relaxation of a test particle in a bath
of field particles. Specifically, we consider a collection of $N$
particles at statistical equilibrium (thermal bath) and introduce a
new particle in the system. To leading order in $N\rightarrow
+\infty$, the particle is advected by the mean flow in phase
space. However, due to finite $N$ effects (graininess), the test
particle undergoes discrete interactions with the particles of the
bath and progressively acquires their distribution. We wish to study
this stochastic process. The probability density $P({\bf r},{\bf
v},t)$ of finding the test particle in ${\bf r}$ with velocity ${\bf
v}$ at time $t$ is governed by a Fokker-Planck equation involving a
term of diffusion and a term of friction. These results are well-known
when the system is spatially homogeneous and memory effects can be
neglected, as in the case of plasma physics. In the present work, we
shall develop a method that allows to treat spatially inhomogeneous
systems and that takes into account non-markovian effects.  Our
approach is also valid if the bath is made of an out-of-equilibrium
distribution of field particles that evolves {\it slowly} so that it
can be assumed stationary on a timescale $N t_D$, which is the typical
relaxation time of the test particle in the bath. This is the case in
particular for one dimensional systems for which the Lenard-Balescu
collision term vanishes at order $O(1/N)$. Therefore, any stable
steady solution of the Vlasov equation does not evolve on a timescale
$N t_D$ \cite{bd,landaud,hb3}.

\subsection{Diffusion coefficient}\label{sec_d}

The increment of the velocity of the test particle between $t-s$ and
$t$ due to the fluctuations of the force is
\begin{eqnarray}
\Delta v^{\mu}=\int_{t-s}^{t}{\cal F}^{\mu}(t')dt'.
\label{d1}
\end{eqnarray}
After standard calculations (see, e.g., Sec. 4.2 of \cite{curious}), the second moment of the velocity increment can be written
\begin{eqnarray}
\left\langle \frac{\Delta v^{\mu}\Delta v^{\nu}}{2s}\right\rangle=\frac{1}{s}\int_{0}^{s}(s+\tau)\langle {\cal F}^{\mu}(t){\cal F}^{\nu}(t-\tau)\rangle d\tau.
\label{d2}
\end{eqnarray}
We shall assume that the correlation function of the force decreases
more rapidly than $\tau^{-1}$ (note, parenthetically, that this is not
the case for the correlation function of the gravitational force which
precisely decreases as $\tau^{-1}$ \cite{ct}). Then, taking the limit
$s\rightarrow +\infty$, we find that the diffusion coefficient is
given by the Kubo formula
\begin{eqnarray}
D^{\mu\nu}=\left\langle \frac{\Delta v^{\mu}\Delta v^{\nu}}{2\Delta t}\right\rangle\equiv \lim_{s\rightarrow +\infty}\left\langle \frac{\Delta v^{\mu}\Delta v^{\nu}}{2s}\right\rangle=\int_{0}^{+\infty}\langle {\cal F}^{\mu}(t){\cal F}^{\nu}(t-\tau)\rangle d\tau.
\label{d3}
\end{eqnarray}
On the other hand, after straightforward calculations (see, e.g.,
Sec. 4.1 of \cite{curious}), we obtain
\begin{eqnarray}
\langle {\cal F}^{\mu}(t){\cal F}^{\nu}(t-\tau)\rangle=N\langle {\cal F}^{\mu}(1\rightarrow 0,t){\cal F}^{\nu}(1\rightarrow 0,t-\tau)\rangle\nonumber\\
=\int d{\bf r}_{1}d{\bf v}_{1}{\cal F}^{\mu}(1\rightarrow 0,t) {\cal
F}^{\nu}(1\rightarrow 0,t-\tau)\frac{f}{m}({\bf r}_{1},{\bf v}_{1}).
\label{d4}
\end{eqnarray}
Therefore, combining Eqs. (\ref{d3}) and (\ref{d4}), we get
\begin{eqnarray}
D^{\mu\nu}=\int_{0}^{+\infty} d\tau d{\bf r}_{1}d{\bf v}_{1}{\cal
F}^{\mu}(1\rightarrow 0,t){\cal F}^{\nu}(1\rightarrow
0,t-\tau)\frac{f}{m}({\bf r}_{1},{\bf v}_{1}). \label{d5}
\end{eqnarray}
For a spatially homogeneous distribution, the diffusion coefficient
reduces to
\begin{eqnarray}
D^{\mu\nu}=\int_{0}^{+\infty} d\tau d{\bf r}_{1}d{\bf
v}_{1}{F}^{\mu}(1\rightarrow 0,t){F}^{\nu}(1\rightarrow
0,t-\tau)\frac{f}{m}({\bf v}_{1}). \label{d6}
\end{eqnarray}
If we neglect collective effects, the force (by unit of mass) created
by the field particle $1$ on the test particle $0$ can be written  
(see Paper I):
\begin{eqnarray}
{\bf F}(1\rightarrow 0,t)=-im\int {\bf k}\hat{u}(k)e^{i{\bf k}({\bf
r}-{\bf r}_{1})}d{\bf k}. \label{d7}
\end{eqnarray}
At time $t-\tau$, we have
\begin{eqnarray}
{\bf F}(1\rightarrow 0,t-\tau)=-im\int {\bf k}\hat{u}(k)e^{i{\bf
k}({\bf r}(t-\tau)-{\bf r}_{1}(t-\tau))}d{\bf k}. \label{d8}
\end{eqnarray}
To leading order in $N\rightarrow +\infty$, the particles
follow rectilinear trajectories so that  ${\bf r}_{i}(t-\tau)={\bf
r}_{i}-{\bf v}_{i}\tau$ where ${\bf r}_i={\bf r}_i(t)$ and ${\bf v}_i={\bf
v}_i(t)$ denote their position and velocity at time $t$. Then, we
get (with ${\bf x}={\bf r}-{\bf r}_{1}$ and ${\bf w}={\bf v}-{\bf v}_{1}$): 
\begin{eqnarray}
{\bf F}(1\rightarrow 0,t-\tau)=-im\int {\bf k}\hat{u}(k)e^{i{\bf k}({\bf x}-{\bf w}\tau)}d{\bf k}.
\label{d9}
\end{eqnarray}
Substituting this expression in Eq. (\ref{d6}) and carrying the
integrations on ${\bf r}_1$ and $\tau$, we obtain after
straightforward calculations
\begin{eqnarray}
D^{\mu\nu}=\pi (2\pi)^{d}m\int k^{\mu}k^{\nu}\hat{u}(k)^{2}\delta({\bf k}\cdot {\bf w})f({\bf v}_{1})d{\bf k}d{\bf v}_{1}.
\label{d10}
\end{eqnarray}
If we take into account collective effects (see Appendix
\ref{sec_coll}), we have to replace Eq. (\ref{d9}) by
\begin{eqnarray}
{\bf F}(1\rightarrow 0,t-\tau)=-im\int {\bf k}\frac{\hat{u}(k)}{\epsilon({\bf k},{\bf k}\cdot {\bf v}_{1})}e^{i{\bf k}({\bf x}-{\bf w}\tau)}d{\bf k}.
\label{d11}
\end{eqnarray}
Then, we get
\begin{eqnarray}
D^{\mu\nu}=\pi (2\pi)^{d}m\int k^{\mu}k^{\nu}\frac{\hat{u}(k)^{2}}{|\epsilon({\bf k},{\bf k}\cdot {\bf v})|^{2}}\delta({\bf k}\cdot {\bf w})f({\bf v}_{1})d{\bf k}d{\bf v}_{1}.
\label{d12}
\end{eqnarray}
The calculation of the diffusion coefficient tensor $D^{\mu\nu}$ for
different potentials of interaction and different dimensions of space
is performed in Paper II and in \cite{landaud}. For one dimensional
systems, we have the simple result
\begin{eqnarray}
D(v)=4\pi^{2}m f(v)\int_{0}^{+\infty}\frac{k\hat{u}(k)^{2}}{|\epsilon({k},{k}{v})|^{2}}dk,
\label{d13}
\end{eqnarray}
where we have used $\delta(k(v-v_{1}))=(1/|k|)\delta(v-v_{1})$ to perform the integration on $v_1$.

\subsection{Friction coefficient}\label{sec_f}

In addition to its diffusive motion, a test particle evolving in a
bath of field particles undergoes a dynamical friction. The friction
corresponds to the response of the field particles to the perturbation
caused by the test particle, as in a polarization process. The test
particle modifies the distribution of the field particles and the
retroaction of this perturbation on the test particle creates a
friction. The expression of the friction force can be derived from a
linear response theory starting from the Liouville equation as done in
Kandrup \cite{kandrup2}. In this section, we show that it can also be
obtained from the Klimontovich equation. This will make a close
connection to the quasilinear theory developed in Sec. \ref{sec_quasi}.

The introduction of a test particle in a bath of field particles
modifies the distribution function $f({\bf r},{\bf v},t)$ of the
bath. Since this perturbation is small, it can be described by the
linearized equation
\begin{eqnarray}
\frac{\partial \delta f}{\partial t}+L\delta f= \nabla\delta\Phi\frac{\partial f}{\partial {\bf v}},
\label{f1}
\end{eqnarray}
whose  formal solution is
\begin{eqnarray}
\delta f(t)=\int_{0}^{t}d\tau G(t,t-\tau)\nabla\delta\Phi (t-\tau)\frac{\partial f}{\partial {\bf v}}(t-\tau).
\label{f2}
\end{eqnarray}
We have used the fact that, initially, $\delta f(0)=0$.  On the other
hand, the perturbation of the force in ${\bf r}$ is given by
\begin{eqnarray}
-\nabla\delta\Phi({\bf r},t)=\frac{1}{m}\int {\bf F}(1\rightarrow 0)\delta f_{1}(t)d{\bf x}_{1}+\int {\bf \cal F}(1\rightarrow 0)\delta({\bf r}_{1}-{\bf r}_{P}(t)) d{\bf r}_{1},
\label{f3}
\end{eqnarray}
where ${\bf r}_{P}(t)$ denotes the position of the test particle.
The second term is the force created by the test particle and the
first term is the fluctuation of the force due to the perturbed density distribution  of the field
particles. Substituting Eq. (\ref{f2}) in Eq. (\ref{f3}) we obtain
\begin{eqnarray}
-\nabla\delta\Phi({\bf r},t)=\frac{1}{m}\int_{0}^{t}d\tau\int d{\bf x}_{1}{\bf F}(1\rightarrow 0)G_{1}(t,t-\tau)\frac{\partial\delta\Phi_{1}}{\partial r_{1}^{\nu}} (t-\tau)\frac{\partial f_{1}}{\partial v_{1}^{\nu}} (t-\tau)\nonumber\\
+\int {\bf \cal F}(1\rightarrow 0)\delta({\bf r}_{1}-{\bf r}_{P}(t)) d{\bf r}_{1}.
\label{f4}
\end{eqnarray}
This is an integral equation for $-\nabla\delta\Phi({\bf r},t)$. For a
spatially homogeneous system, one can solve this equation exactly by
using Laplace-Fourier transforms.  This is how the dielectric function
enters in the problem (see Appendix \ref{sec_coll}). In order to treat
more general systems that are not necessarily homogeneous, we shall
make an approximation which amounts to neglecting some collective
effects.  We solve Eq. (\ref{f4}) by an iterative process: we first
neglect the first term in the r.h.s. of Eq. (\ref{f4}) keeping only
the contribution of the test particle. Then, we substitute this value
in the first term of the r.h.s of Eq. (\ref{f4}). This operation gives
\begin{eqnarray}
-\nabla\delta\Phi({\bf r},t)=-\frac{1}{m}\int_{0}^{t}d\tau\int d{\bf x}_{1}d{\bf r}_{2}{\bf F}(1\rightarrow 0)G_{1}(t,t-\tau){\cal F}^{\nu}(2\rightarrow 1)\nonumber\\
\times\frac{\partial f_{1}}{\partial v_{1}^{\nu}} (t-\tau)\delta({\bf r}_{2}-{\bf r}_{P}(t-\tau))
+\int {\bf \cal F}(1\rightarrow 0)\delta({\bf r}_{1}-{\bf r}_{P}(t))d{\bf r}_{1}.
\label{f5}
\end{eqnarray}
This quantity represents the fluctuation of the field in ${\bf r}$
caused by the introduction of the test particle in the system and
taking into account of the retroaction of the field particles. If we
evaluate this expression at the position ${\bf r}_P$ of the
test particle and subtract the second term (self-interaction),
we obtain the friction force felt by the test particle in response to the
perturbation that it caused. Denoting now by $0$ the position of the test particle, we find that the friction is given by
\begin{eqnarray}
F^{\mu}_{pol}=-\frac{1}{m}\int_{0}^{t}d\tau\int d{\bf
r}_{1}d{\bf v}_{1}{F}^{\mu}(1\rightarrow 0,t){\cal
F}^{\nu}(0\rightarrow 1,t-\tau)\frac{\partial f}{\partial v^{\nu}} ({\bf
r}_{1}(t-\tau),{\bf v}_{1}(t-\tau)). \label{f6}
\end{eqnarray}
For a thermal bath, where the distribution of the field particles is given by  $f({\bf r}_{1},{\bf v}_{1})=A e^{-\beta m (v_{1}^{2}/2+\Phi({\bf r}_{1}))}$, we obtain
\begin{eqnarray}
F^{\mu}_{pol}=\beta\int_{0}^{t}d\tau\int d{\bf
r}_{1}d{\bf v}_{1}{F}^{\mu}(1\rightarrow 0,t){\cal
F}(0\rightarrow 1,t-\tau)\cdot {\bf v}_{1}(t-\tau) f({\bf
r}_{1},{\bf v}_{1}), \label{f6bis}
\end{eqnarray}
where we have used $f({\bf r}_{1}(t-\tau),{\bf v}_{1}(t-\tau))=
f({\bf r}_{1}(t),{\bf v}_{1}(t))$ since $f$ is a stationary solution
of the Vlasov equation. This is equivalent to the result of Kandrup
\cite{kandrup2} based on the Liouville equation but it is obtained
here in a simpler manner from the Klimontovich equation. We can also
obtain this result in a slightly different way. We approximate
$-\nabla\delta\Phi({\bf r},t)$ in Eq. (\ref{f1}) by the force ${\cal
F}(P\rightarrow 0)$ created by the test particle only so that
\begin{eqnarray}
\frac{\partial \delta f}{\partial t}+L\delta f= -{\cal F}(P\rightarrow 0)\frac{\partial f}{\partial {\bf v}}.
\label{fadd1}
\end{eqnarray}
This equation can be solved with a Green function yielding
\begin{eqnarray}
\delta f(t)=-\int_{0}^{t}d\tau G(t,t-\tau){\cal F}(P\rightarrow 0,t-\tau)\frac{\partial f}{\partial {\bf v}}(t-\tau).
\label{fadd2}
\end{eqnarray}
This represents the perturbation of the distribution function of the  field
particles caused by the introduction of a test particle in the
system. This perturbation produces in turn a force which acts as a
friction on the test particle (by retroaction). If we substitute
Eq. (\ref{fadd2}) in the first part of Eq. (\ref{f3}) and evaluate this quantity at the position of the test particle, we recover Eq. (\ref{f6}) for the friction. 

If we now consider a spatially homogeneous distribution of field
particles, the expression of the friction force becomes
\begin{eqnarray}
F^{\mu}_{pol}=\frac{1}{m}\int_{0}^{t}d\tau\int d{\bf
r}_{1}d{\bf v}_{1}{F}^{\mu}(1\rightarrow 0,t){F}^{\nu}(1\rightarrow 0,t-\tau)\frac{\partial f}{\partial v^{\nu}} ({\bf
v}_{1}), \label{f7}
\end{eqnarray}
where we have used ${\bf v}_{1}(t-\tau)={\bf v}_{1}(t)$ to leading
order in $N\rightarrow +\infty$. Taking the limit $t\rightarrow
+\infty$, we get
\begin{eqnarray}
F^{\mu}_{pol}=\frac{1}{m}\int_{0}^{+\infty}d\tau\int d{\bf
r}_{1}d{\bf v}_{1}{F}^{\mu}(1\rightarrow 0,t){F}^{\nu}(1\rightarrow
0,t-\tau)\frac{\partial f}{\partial v^{\nu}} ({\bf v}_{1}). \label{f8}
\end{eqnarray}
This is a sort of generalized Kubo relation involving the gradient of
the distribution function in velocity space instead of the
distribution function itself. {\it The nice similarity in the expressions
of the diffusion coefficient (\ref{d6}) and friction force (\ref{f8})
is worth mentioning.} The integrals on ${\bf r}_{1}$ and $\tau$ can be
calculated in the same manner as in Sec. \ref{sec_d} and we obtain
\begin{eqnarray}
F^{\mu}_{pol}=\pi (2\pi)^{d}m\int d{\bf v}_{1}d{\bf
k}\hat{u}(k)^{2}k^{\mu}k^{\nu}\delta({\bf k}\cdot {\bf
w})\frac{\partial f_{1}}{\partial v_{1}^{\nu}}. \label{f9}
\end{eqnarray}

In order to take into account collective effects, we can follow the
approach of Hubbard \cite{hubbard}. The force (by unit of mass)
created in ${\bf r}$ by the introduction of the test particle is
\begin{eqnarray}
{\bf F}(P\rightarrow 0)=-im\int {\bf k}\frac{\hat{u}(k)}{\epsilon({\bf k},{\bf k}\cdot {\bf v}_{P})}e^{i{\bf k}({\bf r}-{\bf r}_{P})}d{\bf k},
\label{f10}
\end{eqnarray}
where the dielectric function takes into account the response of the
whole system. The bare force due to the test particle alone is
\begin{eqnarray}
{\bf F}(P\rightarrow 0)=-im\int {\bf k}\hat{u}(k)e^{i{\bf k}({\bf r}-{\bf r}_{P})}d{\bf k}.
\label{f11}
\end{eqnarray}
If we subtract Eq. (\ref{f11}) from Eq. (\ref{f10}), we get the force created in ${\bf r}$ by the perturbation of the distribution function of the field particles caused by the introduction of the test particle. Evaluating this force at the position of the test particle, we obtain the friction that it experiences as a result of the polarization process
\begin{eqnarray}
{\bf F}_{pol}=-im\int {\bf k}\hat{u}(k)\left \lbrack
\frac{1}{\epsilon({\bf k},{\bf k}\cdot {\bf v})}-1\right \rbrack
d{\bf k}. \label{f12}
\end{eqnarray}
This  can also be written
\begin{eqnarray}
{\bf F}_{pol}=m\int {\bf k}\hat{u}(k){\rm Im}\left \lbrack
\frac{1}{\epsilon({\bf k},{\bf k}\cdot {\bf v})}\right \rbrack d{\bf
k}. \label{f13}
\end{eqnarray}
Using the identity (B12) of Paper II, we finally obtain
\begin{eqnarray}
F^{\mu}_{pol}=\pi (2\pi)^{d}m\int d{\bf v}_{1}d{\bf
k}\frac{\hat{u}(k)^{2}}{|\epsilon({\bf k},{\bf k}\cdot {\bf
v})|^{2}}k^{\mu}k^{\nu}\delta({\bf k}\cdot {\bf w})\frac{\partial
f_{1}}{\partial v_{1}^{\nu}}. \label{f14}
\end{eqnarray}
If we neglect collective effects and take $|\epsilon({\bf k},{\bf
k}\cdot {\bf v})|^{2}=1$, we recover Eq. (\ref{f9}) obtained in a
different manner.  Now, the friction force is due not only to the
polarization but also to the variation of the diffusion coefficient
with the velocity of the test particle ${\bf v}$. As a result, the
complete expression of the friction force is
\begin{eqnarray}
F^{\mu}_{friction}\equiv \left\langle \frac{\Delta v^{\mu}}{\Delta
t}\right\rangle =F^{\mu}_{pol}+\frac{\partial D^{\mu\nu}}{\partial
v^{\nu}}. \label{f15}
\end{eqnarray}
The second term is obtained when we take into account the influence of
the fluctuations of the force in the trajectory of the test particle,
i.e. when we go beyond the rectilinear trajectory approximation. As
shown by Hubbard
\cite{hubbard}, this is necessary for the calculation of the friction while this is not necessary for the calculation of the diffusion coefficient. From Eqs. (\ref{d12}) and (\ref{f14}) we get
\begin{eqnarray}
F^{\mu}_{friction}=\pi (2\pi)^{d}m\int d{\bf v}_{1}d{\bf
k}k^{\mu}k^{\nu}\hat{u}(k)^{2}f_{1}\left (\frac{\partial}{\partial
v^{\nu}}-\frac{\partial}{\partial v_{1}^{\nu}}\right
)\frac{\delta({\bf k}\cdot {\bf w})}{|\epsilon({\bf k},{\bf k}\cdot
{\bf v})|^{2}},
\label{f16}
\end{eqnarray}
where we have used an integration by parts in Eq. (\ref{f14}). When we
ignore collective effects, expressions (\ref{d12}) and (\ref{f14}) for
the diffusion coefficient and the friction force can be obtained directly
from the Hamiltonian equations, by making a systematic expansion of
the trajectory of the particles in powers of $1/N$ in the limit
$N\rightarrow +\infty$ as shown in Appendix \ref{sec_fs}.

For a thermal bath, corresponding to the case where the field
particles are at statistical equilibrium, the distribution function
is the Maxwell-Boltzmann distribution
\begin{eqnarray}
f({\bf v}_{1})=\left (\frac{\beta m}{2\pi}\right )^{d/2} \rho e^{-\beta m \frac{v_{1}^{2}}{2}}.
\label{f17}
\end{eqnarray}
Inserting the identity
\begin{eqnarray}
\frac{\partial f}{\partial{\bf v}_1}=-\beta m f_1 {\bf
v}_1,\label{f18}
\end{eqnarray}
in Eq. (\ref{f14}), using the $\delta$-function to replace ${\bf
k}\cdot {\bf v}_1$ by ${\bf k}\cdot {\bf v}$, and comparing with
Eq. (\ref{d12}), we find that
\begin{eqnarray}
F^{\mu}_{pol}=-\beta m D^{\mu\nu}{v}^{\nu}. \label{f19}
\end{eqnarray}
This can be viewed as a generalized Einstein relation. We note that
the diffusion coefficient and the friction coefficient depend on the
velocity of the test particle. {\it We also note that the Einstein
relation is valid for the friction force ${\bf F}_{pol}$ due to the
polarization, not for the total friction force (\ref{f16})}. We do not
have this subtlety for the ordinary Brownian motion where the
diffusion coefficient is constant. 

We now consider an arbitrary (steady) distribution of the bath. If we
neglect collective effects and use Eqs. (\ref{d10}) and (\ref{f9}) we
obtain after simple manipulations (see Eq. (16) in \cite{landaud}):
\begin{eqnarray}
\frac{\partial D^{\mu\nu}}{\partial v^{\nu}}=F^{\mu}_{pol}.
\label{f20}
\end{eqnarray}
Therefore,
\begin{eqnarray}
{\bf F}_{friction}=2{\bf F}_{pol}. \label{f21}
\end{eqnarray}
We note that the friction force calculated by Kandrup \cite{kandrup2}
corresponds to the polarization part ${\bf F}_{pol}$ while
Chandrasekhar \cite{kc} computes the full friction ${\bf
F}_{friction}$. This explains why there is a factor $1/2$ between
their results for equal mass particles (see \cite{kandrup2},
pp. 446). 

Finally, for 1D systems, we have the simple result
\begin{eqnarray}
F_{pol}=4\pi^{2}m
f'(v)\int_{0}^{+\infty}\frac{k\hat{u}(k)^{2}}{|\epsilon({k},{k}{v})|^{2}}dk.
\label{f22}
\end{eqnarray}
This expression is valid for an arbitrary (steady) distribution of the bath and
it takes into account collective effects. Comparing Eq. (\ref{f22}) with
Eq. (\ref{d13}), we find that the friction force is related to the
diffusion coefficient by the relation
\begin{eqnarray}
F_{pol}=D(v)\frac{d\ln f}{dv}. \label{f23}
\end{eqnarray}
This can be viewed as a generalization of the Einstein relation for an out-of-equilibrium distribution of the bath.

\subsection{The Fokker-Planck equation}\label{sec_p}

Assuming that the system is spatially homogeneous, the 
probability density $P({\bf v},t)$ of finding the test particle with the
velocity ${\bf v}$ at time $t$ is governed by a  Fokker-Planck
equation of the form
\begin{equation}
\label{p1} {\partial P\over\partial t}={1\over 2}{\partial^{2}\over\partial v^{\mu}\partial v^{\nu}}\biggl (P{\langle \Delta v^{\mu} \Delta v^{\nu}\rangle\over \Delta t}\biggr )-{\partial\over\partial v^{\mu}}\biggl (P{\langle \Delta v^{\mu}\rangle\over \Delta t}\biggr ).
\end{equation}
This Fokker-Planck approach assumes that the stochastic process is
markovian (see Sec. \ref{sec_nm} for generalizations). It also assumes
that the higher order moments of the increment of velocity $\Delta v$
play a negligible role. This is indeed the case in the $N\rightarrow
+\infty$ limit that we consider since they are of order $O(N^{-2})$ or
smaller. At order $O(N^{-1})$, we have found that the second
(diffusion) and first (friction) moments of the velocity increment of
the test particle are given by
\begin{equation}
\label{p2}{\langle \Delta v^{\mu} \Delta v^{\nu}\rangle\over 2\Delta t}=D^{\mu\nu}, \qquad {\langle \Delta v^{\mu}\rangle\over \Delta t}={\partial D^{\mu\nu}\over\partial v^{\nu}}+\eta^{\mu},
\end{equation}
with
\begin{eqnarray}
D^{\mu\nu}=\pi (2\pi)^{d}m\int 
k^{\mu}k^{\nu}\frac{\hat{u}(k)^{2}}{|\epsilon({\bf k},{\bf k}\cdot
{\bf v})|^{2}}\delta({\bf k}\cdot {\bf w})f({\bf v}_{1})d{\bf k}d{\bf v}_{1},
\label{p3}
\end{eqnarray}
\begin{eqnarray}
\eta^{\mu}\equiv F^{\mu}_{pol}=\pi (2\pi)^{d}m\int \frac{\hat{u}(k)^{2}}{|\epsilon({\bf k},{\bf k}\cdot {\bf
v})|^{2}}k^{\mu}k^{\nu}\delta({\bf k}\cdot {\bf w})\frac{\partial
f_{1}}{\partial v_{1}^{\nu}}d{\bf v}_{1}d{\bf
k}. \label{p4}
\end{eqnarray}
Note that we have changed the sign of $\eta^{\mu}$ with respect to
Paper II. The Fokker-Planck equation (\ref{p1}) can be written in the
alternative form
\begin{equation}
\label{p5}{\partial P\over\partial t}={\partial\over\partial v^{\mu}}\biggl (D^{\mu\nu}{\partial P\over\partial v^{\nu}}-P\eta^{\mu}\biggr ).
\end{equation}
The two expressions (\ref{p1}) and (\ref{p5}) have their own interest. The
expression (\ref{p1}) where the diffusion coefficient is placed after the
second derivative $\partial^{2}(DP)$ involves the total friction
force $F_{friction}^{\mu}=\langle \Delta v^{\mu}\rangle/\Delta t$
and the expression (\ref{p5}) where the diffusion coefficient is placed
between the derivatives $\partial D\partial P$ isolates the part of
the friction $\eta^{\mu}= F^{\mu}_{pol}$ due to the polarization.
This alternative form (\ref{p5})  has therefore a clear physical
interpretation. Inserting the expressions (\ref{p3}) and (\ref{p4}) of the
diffusion coefficient and friction term in Eq. (\ref{p5}), we obtain
\begin{eqnarray}
\label{p6} {\partial P\over\partial t}=\pi (2\pi)^{d}m{\partial\over\partial v^{\mu}}\int d{\bf v}_{1}d{\bf k}k^{\mu}k^{\nu}\frac{\hat{u}({k})^{2}}{|\epsilon({\bf k},{\bf k}\cdot {\bf v})|^{2}}\delta\lbrack {\bf k}\cdot ({\bf v}-{\bf v}_{1})\rbrack \biggl ({\partial \over\partial v^{\nu}}-{\partial \over\partial {v}_{1}^{\nu}}\biggr )f({\bf v}_{1})P({\bf v},t).
\end{eqnarray}
For a thermal bath, using Eqs. (\ref{f19}), the Fokker-Planck equation
(\ref{p5}) can be written
\begin{equation}
\label{p7}{\partial P\over\partial t}={\partial\over\partial v^{\mu}}\biggl \lbrack D^{\mu\nu}\biggl ({\partial P\over\partial v^{\nu}}+\beta m P v^{\nu}\biggr )\biggr \rbrack,
\end{equation}
where $D^{\mu\nu}(v)$ is given by Eq. (\ref{p3}). Since the r.h.s. of
Eq. (\ref{p7}) is of order $O(1/N)$, the distribution of the test
particle $P({\bf v},t)$ relaxes to the Maxwellian distribution on a
typical timescale $Nt_D$ (see \cite{landaud} for more details).  In
one dimension, the bath $f(v)$ can be any stable stationary solution of the Vlasov equation. Using
Eq. (\ref{f23}), the Fokker-Planck equation (\ref{p5}) can be written
\begin{equation}
\label{p8}{\partial P\over\partial t}={\partial\over\partial v}\biggl \lbrack D\biggl ({\partial P\over\partial v}-P \frac{d\ln f}{dv}\biggr )\biggr \rbrack,
\end{equation}
where $D(v)$ is given by Eq. (\ref{d13}). The distribution of the test
particle $P({v},t)$ relaxes to the distribution of the bath $f(v)$ on a
typical timescale $Nt_D$ \cite{landaud}.

In Paper II, we have obtained the Fokker-Planck equation (\ref{p6})
from the Lenard-Balescu equation (II-49) by replacing $f({\bf v},t)$
by the distribution of the test particle $P({\bf v},t)$ and $f({\bf
v}_1,t)$ by the static distribution of the bath $f({\bf v}_1)$. This
procedure transforms an integrodifferential equation (II-49) in a
differential equation (\ref{p6}). The expressions
(\ref{p2})-(\ref{p4}) of the diffusion and friction were then obtained
by identifying Eq. (\ref{p6}) with the Fokker-Planck equation
(\ref{p1}). In the present paper, we have proceeded the other way
round by first determining the moments (\ref{p2})-(\ref{p4}), then
inserting them in the Fokker-Planck equation (\ref{p1}). Note that
Hubbard \cite{hubbard} derived the expressions (\ref{p2})-(\ref{p4})
of the diffusion coefficient and friction force but did not make the
calculations explicitly until the end. In particular, he did not
explicitly wrote down the kinetic equation (\ref{p6}) that is related
to the Lenard-Balescu equation (II-49) discovered independently at the
same period \cite{lenard,balescu}.

\subsection{The Fokker-Planck equation at $T=0$}\label{sec_pz}

In Paper II and in \cite{landaud}, we have given various expressions
of the Fokker-Planck equation (\ref{p6}) for different potentials of
interaction and different dimensions of space. However, we have not
explicitly considered the case $T=0$ which presents interesting
features. At $T=0$, the Maxwell-Boltzmann distribution (\ref{f17}) reduces to
$f({\bf v}_{1})=\rho\delta({\bf v}_{1})$. Substituting this expression in Eq. (\ref{p3}), we find that the diffusion coefficient becomes
\begin{eqnarray}
D^{\mu\nu}=\pi (2\pi)^{d}\rho m\int k^{\mu}k^{\nu}\frac{\hat{u}(k)^{2}}{|\epsilon({\bf k},0)|^{2}}\delta({\bf k}\cdot {\bf v})d{\bf k},
\label{p9}
\end{eqnarray}
with $\epsilon({\bf k},0)=1+(2\pi)^{d}\hat{u}(k)\beta\rho m$ according
to Eq. (II-13). Note that, in this section, we consider the case of
repulsive potentials with $\hat{u}(k)>0$ so that the homogeneous phase
is stable even at $T=0$ (see Paper I). We now observe that the
integral in Eq. (\ref{p9}) is similar to the one in
Eq. (II-41). Therefore, it can be written
\begin{eqnarray}
D^{\mu\nu}=\frac{K_{d}}{v}\left (\delta^{\mu\nu}-\frac{v^{\mu}v^{\nu}}{v^{2}}\right ),
\label{p10}
\end{eqnarray} 
where 
\begin{eqnarray}
 K_{d}=\lambda_{d}\rho m\int_{0}^{+\infty}k^{d}\left\lbrack \frac{\hat{u}(k)}{1+(2\pi)^{d}\hat{u}(k)\beta\rho m}\right\rbrack^{2}dk,
\label{p11}
\end{eqnarray} 
with $\lambda_{3}=8\pi^{5}$ and $\lambda_{2}=8\pi^{3}$. For the
Coulombian potential, we have $(2\pi)^{3}\hat{u}(k)\beta\rho
m=k_{D}^{2}/k^{2}$ where $k_{D}$ is the Debye wavenumber (see Paper
I). Therefore, the collective effects encapsulated in the dielectric
function in the denominator of Eq. (\ref{p11}) regularise the integral
for $k\rightarrow 0$ (this is a particular case of the Lenard-Balescu
equation). On the other hand, noting that $D^{\mu\nu}v^{\nu}=0$
according to Eq. (\ref{p10}), we find that the friction force
(\ref{f19}) vanishes. Therefore, at $T=0$, the Fokker-Planck equation
(\ref{p6}) can be written
\begin{eqnarray}
\frac{\partial P}{\partial t}=K_{d}\frac{\partial}{\partial v^{\mu}}\left ( \frac{\delta^{\mu\nu}v^{2}-v^{\mu}v^{\nu}}{v^{3}}\frac{\partial P}{\partial v^{\nu}}\right ).
\label{p12}
\end{eqnarray} 
This equation admits an infinity of stationary solutions. Indeed,
since $D^{\mu\nu}v^{\nu}=0$, any distribution $P=P(v)$ depending only
on the modulus $v=|{\bf v}|$ of the velocity is a stationary solution
of Eq. (\ref{p12}). Therefore, at $T=0$, the test particle does not
necessarily relax to the distribution of the bath $f({\bf v})=\rho
\delta({\bf v})$. On the other hand, for one dimensional systems,
Eq. (\ref{p12}) reduces to $\partial P/\partial t=0$ so that the
distribution of the test particle does not evolve in time.

\subsection{More general kinetic equations}\label{sec_mg}

It is instructive to compare the Fokker-Planck equation (\ref{p6})
with the more general equation obtained from the projection operator
formalism \cite{kandrup1}. When collective effects are ignored, this
equation can be written
\begin{eqnarray}
\frac{\partial P}{\partial t}+{\bf v}\frac{\partial P}{\partial {\bf r}}+\langle {\bf F}\rangle \frac{\partial P}{\partial {\bf v}}=\frac{\partial}{\partial v^{\mu}}\int_{0}^{t}d\tau\int d{\bf r}_{1}d{\bf v}_{1}{F}^{\mu}(1\rightarrow 0)
G(t,t-\tau)\nonumber\\
\times \biggl\lbrace {\cal F}^{\nu}(1\rightarrow 0)\frac{\partial}{\partial v^{\nu}}
+ {\cal F}^{\nu}(0\rightarrow 1)\frac{\partial}{\partial v_{1}^{\nu}}\biggr\rbrace {P}({\bf r},{\bf v},t-\tau)\frac{f}{m}({\bf r}_{1},{\bf v}_{1}).
\label{mg1}
\end{eqnarray}
It can be obtained from Eq. (\ref{c13}) by replacing $f({\bf v},t)$
by $P({\bf v},t)$ and $f({\bf v}_1,t)$ by $f({\bf v}_1)$. This is a
sort of generalized ``Fokker-Planck'' equation involving a term of
``diffusion'' and a term of ``friction''. However, strictly speaking,
Eq. (\ref{mg1}) is not a Fokker-Planck equation because it is
non-Markovian. We also note that the ``diffusion'' term appears as a
complicated time integral of the force correlation function involving
${P}({\bf r},{\bf v},t-\tau)$. This can be seen as a generalization of
the Kubo formula (\ref{d5}). Similarly the ``friction'' force is a
generalization of the expression obtained in Eq. (\ref{f6}) with a
more complicated time integral. If we consider a thermal bath where the distribution of the field particles is the Boltzmann distribution, we get
\begin{eqnarray}
\frac{\partial P}{\partial t}+{\bf v}\frac{\partial P}{\partial {\bf r}}+\langle {\bf F}\rangle \frac{\partial P}{\partial {\bf v}}=\frac{\partial}{\partial v^{\mu}}\int_{0}^{t}d\tau\int d{\bf r}_{1}d{\bf v}_{1}{F}^{\mu}(1\rightarrow 0)
G(t,t-\tau)\nonumber\\
\times \biggl\lbrace {\cal F}(1\rightarrow 0)\cdot \frac{\partial}{\partial {\bf v}}
-\beta m {\cal F}(0\rightarrow 1)\cdot {\bf v}_{1}\biggr\rbrace {P}({\bf r},{\bf v},t-\tau)\frac{f}{m}({\bf r}_{1},{\bf v}_{1}).
\label{mg1bis}
\end{eqnarray}
If we come back to Eq. (\ref{mg1}), make a Markovian approximation and extend the time integration to infinity, we get 
\begin{eqnarray}
\frac{\partial P}{\partial t}+{\bf v}\frac{\partial P}{\partial {\bf r}}+\langle {\bf F}\rangle \frac{\partial P}{\partial {\bf v}}=\frac{\partial}{\partial v^{\mu}}\int_{0}^{+\infty}d\tau\int d{\bf r}_{1}d{\bf v}_{1}{F}^{\mu}(1\rightarrow 0)
G(t,t-\tau)\nonumber\\
\times \biggl\lbrace {\cal F}^{\nu}(1\rightarrow 0)\frac{\partial}{\partial v^{\nu}}
+ {\cal F}^{\nu}(0\rightarrow 1)\frac{\partial}{\partial v_{1}^{\nu}}\biggr\rbrace {P}({\bf r},{\bf v},t)\frac{f}{m}({\bf r}_{1},{\bf v}_{1}),
\label{mg1b}
\end{eqnarray}
where we recall that the coordinates appearing after the Greenian must
be viewed as explicit functions of time ${\bf r}_{i}(t-\tau)$ and ${\bf
v}_{i}(t-\tau)$ (see Paper III for more details). For a spatially
homogeneous system, Eq. (\ref{mg1}) takes the simplest form
\begin{eqnarray}
\frac{\partial P}{\partial t}=\frac{\partial}{\partial v^{\mu}}\int_{0}^{t}d\tau\int d{\bf r}_{1}d{\bf v}_{1}{F}^{\mu}(1\rightarrow 0,t)
{F}^{\nu}(1\rightarrow 0,t-\tau)\biggl 
(\frac{\partial}{\partial v^{\nu}}
-\frac{\partial}{\partial v_{1}^{\nu}}\biggr ) {P}({\bf
v},t-\tau)\frac{f}{m}({\bf v}_{1}),
\label{mg2}
\end{eqnarray}
where we have used ${\bf v}_{i}(t-\tau)={\bf v}_{i}$ for a spatially
homogeneous system.  We shall come back to this non-markovian equation
in Sec. \ref{sec_nm}. If we now make a Markovian approximation
${P}({\bf v},t-\tau)\simeq {P}({\bf v},t)$ and extend the time
integral to infinity, we get
\begin{eqnarray}
\frac{\partial P}{\partial t}=\frac{\partial}{\partial v^{\mu}}\int_{0}^{+\infty}d\tau\int d{\bf r}_{1}d{\bf v}_{1}{F}^{\mu}(1\rightarrow 0,t)
{F}^{\nu}(1\rightarrow 0,t-\tau) \biggl
(\frac{\partial}{\partial v^{\nu}} - \frac{\partial}{\partial
v_{1}^{\nu}}\biggr ) {P}({\bf v},t)\frac{f}{m}({\bf v}_{1}).
\label{mg3}
\end{eqnarray}
This is a Fokker-Planck equation which can be put in the form (\ref{p5})
with a diffusion coefficient
\begin{eqnarray}
D^{\mu\nu}=\int_{0}^{+\infty} d\tau \int d{\bf r}_{1}d{\bf v}_{1}{F}^{\mu}(1\rightarrow 0,t) {F}^{\nu}(1\rightarrow 0,t-\tau)\frac{f}{m}({\bf v}_{1}),
\label{mg4}
\end{eqnarray}
and a friction force due to the polarization
\begin{eqnarray}
\eta^{\mu}=-\frac{1}{m}\int_{0}^{+\infty}d\tau\int d{\bf r}_{1}d{\bf v}_{1}{F}^{\mu}(1\rightarrow 0,t){F}^{\nu}(0\rightarrow 1,t-\tau)\frac{\partial f}{\partial v^{\nu}} ({\bf v}_{1}).
\label{mg5}
\end{eqnarray}
These expressions agree with Eqs. (\ref{d6}) and (\ref{f8}) obtained
directly from the equations of motion. After integration on $\tau$ and
${\bf r}_1$, we recover the Fokker-Planck equation (\ref{p6}) with the
expressions (\ref{p3}) and (\ref{p4}) of the diffusion coefficient and
friction term (with $|\epsilon({\bf k},{\bf k}\cdot {\bf v})|^{2}=1$
since collective effects are neglected here).

\section{The non-markovian equation}\label{sec_nm}

\subsection{ General results}\label{sec_r}

In this section, we study in more detail the non-Markovian equation
(\ref{mg2}). If the field particles are at statistical equilibrium
(thermal bath), using the identity (\ref{f18}), the non-markovian
equation (\ref{mg2}) takes the form
\begin{eqnarray}
\frac{\partial P}{\partial t}=\frac{\partial}{\partial v^{\mu}}\int_{0}^{t}d\tau\int d{\bf r}_{1}d{\bf v}_{1}{F}^{\mu}(1\rightarrow 0,t)
{F}^{\nu}(1\rightarrow 0,t-\tau)\frac{f}{m}({\bf v}_{1})\biggl (\frac{\partial}{\partial v^{\nu}}
+\beta m v_{1}^{\nu}\biggr ) {P}({\bf v},t-\tau).\nonumber\\
\label{r1}
\end{eqnarray}
It can be rewritten
\begin{eqnarray}
\frac{\partial P}{\partial t}=\frac{\partial}{\partial v^{\mu}}\int_{0}^{t}d\tau\biggl (C^{\mu\nu}(\tau) \frac{\partial}{\partial v^{\nu}}
+\beta m W^{\mu}(\tau)  \biggr ) {P}({\bf v},t-\tau),
\label{r2}
\end{eqnarray}
where we have introduced the notations
\begin{eqnarray}
C^{\mu\nu}(\tau)=\langle F^{\mu}(t)F^{\nu}(t-\tau)\rangle= N\langle F^{\mu}(1\rightarrow 0,t)F^{\nu}(1\rightarrow 0,t-\tau)\rangle,
\label{r3}
\end{eqnarray}
\begin{eqnarray}
W^{\mu}(\tau)=N\langle F^{\mu}(1\rightarrow 0,t)F^{\nu}(1\rightarrow 0,t-\tau)v_{1}^{\nu}\rangle.
\label{r4}
\end{eqnarray}
These quantities can be calculated by making the linear
trajectory approximation. The first quantity has already been
studied in Paper II. It represents the temporal correlation of the
force acting on the test particle. It can be written
\begin{eqnarray}
C^{\mu\nu}(\tau)= (2\pi)^{d} m\int k^{\mu}k^{\nu} \hat{u}(k)^{2}e^{-i{\bf k}({\bf v}-{\bf v}_{1})\tau} f({\bf v}_{1})d{\bf v}_{1}d{\bf k}.
\label{r5}
\end{eqnarray}
Performing the integration on ${\bf v}_{1}$, we get
\begin{eqnarray}
C^{\mu\nu}(\tau)= (2\pi)^{2d} m\int k^{\mu}k^{\nu} \hat{u}(k)^{2}e^{-i{\bf k}{\bf v}\tau} \hat{f}({\bf k}\tau)d{\bf k},
\label{r6}
\end{eqnarray}
where $\hat{f}$ is the Fourier transform of $f$. For a Maxwellian
distribution of the field particles (thermal bath), we have
\begin{eqnarray}
C^{\mu\nu}(\tau)= (2\pi)^{d}\rho m\int k^{\mu}k^{\nu} \hat{u}(k)^{2}e^{-i{\bf k}{\bf v}\tau} e^{-k^{2}\tau^{2}/2\beta m} d{\bf k}.
\label{r7}
\end{eqnarray}
On the other hand, the function ${\bf W}(\tau)$ is given by
\begin{eqnarray}
W^{\mu}(\tau)=m(2\pi)^{d}\int ({\bf k}\cdot {\bf v}_{1}) k^{\mu}\hat{u}(k)^{2}e^{-i{\bf k}({\bf v}-{\bf v}_{1})\tau} f({\bf v}_{1})d{\bf v}_{1}d{\bf k}.
\label{r8}
\end{eqnarray}
Performing the  integration on ${\bf v}_{1}$, we get
\begin{eqnarray}
W^{\mu}(\tau)=-i m(2\pi)^{2d}\int d{\bf k} k^{\mu} \hat{u}(k)^{2}e^{-i{\bf k}{\bf v}\tau} \frac{\partial}{\partial \tau} \hat{f}({\bf k}\tau).
\label{r9}
\end{eqnarray}
For a Maxwellian distribution of the field particles, we obtain
\begin{eqnarray}
W^{\mu}(\tau)=-i \rho m(2\pi)^{d}\int d{\bf k} k^{\mu} \hat{u}(k)^{2}e^{-i{\bf k}{\bf v}\tau} \frac{\partial}{\partial \tau} e^{-k^{2}\tau^{2}/2\beta m},
\label{r10}
\end{eqnarray}
so that, finally,
\begin{eqnarray}
W^{\mu}(\tau)=i  (2\pi)^{d}\rho\frac{\tau}{\beta}\int k^{\mu} \hat{u}(k)^{2}k^{2}e^{-i{\bf k}{\bf v}\tau} e^{-k^{2}\tau^{2}/2\beta m} d{\bf k}.
\label{r11}
\end{eqnarray}
Let us now apply these general results to some specific systems.

\subsection{Self-gravitating systems}\label{sec_a}

For the gravitational interaction, we can easily perform the
integrations in Eqs. (\ref{r7}) and (\ref{r11}) by introducing a
spherical system of coordinates with the $z$ axis in the direction of
${\bf v}$.  The correlation function $C^{\mu\nu}(\tau)$ is given by
Eqs.  (II-94), (II-95) and (II-96). On the other hand, after some
calculations, we find that
\begin{eqnarray}
{\bf W}(\tau)=\frac{4\pi \rho m G^{2}}{v\tau} G(x){\bf v},
\label{a1}
\end{eqnarray}
where ${\bf x}=(\beta m/2)^{1/2}{\bf v}$ and $G(x)$ is the function
defined by Eq. (II-75). Comparing this expression with Eq. (II-95),
we find that
\begin{eqnarray}
{\bf W}(\tau)=C_{\|}(v,\tau){\bf v}=C^{\mu\nu}(v,\tau) v^{\nu}.
\label{a2}
\end{eqnarray}
Therefore, for the gravitational interaction, we have the equality
\begin{eqnarray}
\langle F^{\mu}(1\rightarrow 0,t)F^{\nu}(1\rightarrow 0,t-\tau)v_{1}^{\nu}\rangle =\langle F^{\mu}(1\rightarrow 0,t)F^{\nu}(1\rightarrow 0,t-\tau)\rangle v^{\nu}.
\label{a3}
\end{eqnarray}
We stress, however, that this equality is not true for any
potential. Using the relation (\ref{a2}), we can rewrite the non-Markovian
equation (\ref{r2}) in the form
\begin{eqnarray}
\frac{\partial P}{\partial t}=\frac{\partial}{\partial v^{\mu}}\int_{0}^{t}d\tau C^{\mu\nu}(\tau)\biggl (\frac{\partial}{\partial v^{\nu}}
+\beta m v^{\nu}  \biggr ) {P}({\bf v},t-\tau).
\label{a4}
\end{eqnarray}
For a spherically symmetric system, the distribution $P({\bf v},t)$
depends only on the modulus $|{\bf v}|=v$ of the velocity and we
obtain
\begin{eqnarray}
\frac{\partial P}{\partial t}=\frac{1}{v^{2}}\frac{\partial}{\partial v}\left\lbrack v^{2}\int_{0}^{t}d\tau C_{\|}(\tau,v)\biggl (\frac{\partial}{\partial v}
+\beta m v  \biggr ) {P}({v},t-\tau)\right\rbrack,
\label{a5}
\end{eqnarray}
where (see Paper II):
\begin{eqnarray}
C_{\|}(\tau,v)=\frac{4\pi \rho m G^{2}}{v\tau} G(x).
\label{a6}
\end{eqnarray}
If we make a markovian approximation ${P}({v},t-\tau)\simeq P(v,t)$
and extend the time integration to $+\infty$, we recover the
Kramers-Chandrasekhar equation \cite{kc}:
\begin{eqnarray}
\frac{\partial P}{\partial t}=\frac{1}{v^{2}}\frac{\partial}{\partial v}\left\lbrack v^{2} D_{\|}(v)\biggl (\frac{\partial P}{\partial v}
+\beta m P v  \biggr ) \right\rbrack,
\label{a7}
\end{eqnarray}
with
\begin{eqnarray}
D_{\|}(v)= \int_{0}^{+\infty} C_{\|}(\tau,v)d\tau=\frac{4\pi \rho m G^{2}}{v} G(x) \int_{0}^{+\infty}\frac{d\tau}{\tau}.
\label{a8}
\end{eqnarray}
This expression exhibits the well-known logarithmic divergence of the
diffusion coefficient which appears here in the time integration (see
a discussion of this issue in Paper II). The divergence for
$t\rightarrow 0$ is related to the linear trajectory
approximation and could be cured by a more accurate treatment of
binary collisions. The divergence for $t\rightarrow +\infty$ is more
serious.  One usually introduces a cut-off but this procedure is
relatively {\it ad hoc}.  Alternatively, one could consider the
non-Markovian equation (\ref{a5})-(\ref{a6}) which is well-posed for
any time $t$.

\subsection{The HMF model}\label{sec_m}

As discussed previously, non-markovian effects can be important for
self-gravitating systems because the temporal correlation function of
the force decreases algebraically, like $t^{-1}$. For neutral plasmas,
the situation is different because of Debye shielding. In that case,
collective effects cannot be ignored in the computation of the force
auto-correlation function and they are taken into account through the
dielectric function in Eq. (II-98). Then, the temporal correlation
function is given by Eqs. (II-100), (II-109), (II-110) and (II-21) of
Paper II, and it decreases exponentially rapidly. In that case, the
Markovian approximation is valid. Collective effects are also
important for the HMF model \cite{bouchet,cvb,hb2}.  When collective
effects are ignored, it is found that the temporal decay of the
correlation function of the force is gaussian, see Eq. (II-113). By
contrast, when collective effects are properly accounted for, it is
found that the correlation function decreases exponentially rapidly,
see Eq. (II-111). Furthermore, the decay rate tends to zero for
$T\rightarrow T_{c}$ implying a slow decay of the correlations. This
may unveil a failure of the markovian approximation close to the
critical temperature. For that reason, it may be useful to derive
non-markovian kinetic equations which take into account collective
effects.

Collective effects can be taken into account in the
non-markovian equation (\ref{r2}) by making the substitution
\begin{eqnarray}
\hat{u}(k)^{2}\rightarrow \frac{\hat{u}(k)^{2}}{|\epsilon({\bf k},{\bf k}\cdot {\bf v}_{1})|^{2}},
\label{m1}
\end{eqnarray}
in the expressions (\ref{r5}) and (\ref{r8}). The correlation function of the
force is now given by
\begin{eqnarray}
C^{\mu\nu}(\tau)= (2\pi)^{d} m\int k^{\mu}k^{\nu} \frac{\hat{u}(k)^{2}}{|\epsilon({\bf k},{\bf k}\cdot {\bf v}_{1})|^{2}} e^{-i{\bf k}({\bf v}-{\bf v}_{1})\tau} f({\bf v}_{1})d{\bf v}_{1}d{\bf k}.
\label{m2}
\end{eqnarray}
This can be written
\begin{eqnarray}
C^{\mu\nu}(\tau)= (2\pi)^{d} m\int k^{\mu}k^{\nu} \hat{u}(k)^{2}e^{-i{\bf k}{\bf v}\tau} Q({\bf k},\tau)d{\bf k},
\label{m3}
\end{eqnarray}
where the function $Q({\bf k},\tau)$ is defined by Eq. (II-99). On the
other hand, the function ${\bf W}(\tau)$ is given by
\begin{eqnarray}
W^{\mu}(\tau)=m(2\pi)^{d}\int ({\bf k}\cdot {\bf v}_{1}) k^{\mu}\frac{\hat{u}(k)^{2}}{|\epsilon({\bf k},{\bf k}\cdot {\bf v}_{1})|^{2}} e^{-i{\bf k}({\bf v}-{\bf v}_{1})\tau} f({\bf v}_{1})d{\bf v}_{1}d{\bf k}.
\label{m4}
\end{eqnarray}
This can be rewritten
\begin{eqnarray}
W^{\mu}(\tau)=-i m(2\pi)^{d}\int d{\bf k} k^{\mu} \hat{u}(k)^{2}e^{-i{\bf k}{\bf v}\tau} \frac{\partial}{\partial \tau} Q({\bf k},\tau).
\label{m5}
\end{eqnarray}
For a Maxwellian distribution of the field particles, the large time
asymptotics of $Q({\bf k},\tau)$ is given by (II-109). Using
Eqs. (\ref{m3}) and (\ref{m5}), we can then obtain the large time
asymptotic of $C^{\mu\nu}(\tau)$ and $W^{\mu}(\tau)$ for
$\tau\rightarrow +\infty$. 

Let us now specifically consider the HMF model where the potential of
interaction is truncated to one Fourier mode. For this system, using
Eqs.  (\ref{m3}) and (\ref{m5}), the non-markovian equation (\ref{r2})
can be written
\begin{eqnarray}
\frac{\partial P}{\partial t}=\frac{k^{2}}{4\pi}\frac{\partial}{\partial v}\int_{0}^{t}d\tau \left\lbrack Q(\tau)\cos(v\tau)\frac{\partial}{\partial v}-\beta Q'(\tau)\sin(v\tau)\right\rbrack  {P}({v},t-\tau),
\label{m6}
\end{eqnarray}
where $Q(\tau)$ behaves like
\begin{eqnarray}
Q(\tau)\sim \rho\left (\frac{2}{\beta}\right )^{1/2}\frac{1}{\gamma |F'(\gamma \sqrt{\beta/2})|}e^{-\gamma \tau},
\label{m7}
\end{eqnarray}
for $\tau\rightarrow +\infty$. The damping rate $\gamma$ and the
function $F(x)$ are defined in Paper II. As discussed above, the
exponential relaxation time $\gamma^{-1}(T)$ diverges for
$T\rightarrow T_{c}$ so that the Markovian approximation may not be
correct close to the critical point. This may be an interesting
situation to analyze in deeper detail with the non-markovian
equation (\ref{m6}).

If we neglect collective effects, we find that
\begin{eqnarray}
Q({\bf k},\tau)=(2\pi)^{d}\hat{f}({\bf k}t)=\rho e^{-k^{2}\tau^{2}/2\beta m},
\label{m8}
\end{eqnarray}
where the second equality is valid for a Maxwellian distribution of
the field particles. For the HMF model, we obtain
\begin{eqnarray}
Q(\tau)=\rho e^{-\tau^{2}/2\beta}.
\label{m9}
\end{eqnarray}
This yields a Gaussian decay of the correlations instead of an
exponential decay in Eq. (\ref{m7}) when collective effects are
accounted for \cite{cvb}. With the expression (\ref{m7}) for
$Q(\tau)$, the non-markovian equation (\ref{r2}) becomes
\begin{eqnarray}
\frac{\partial P}{\partial t}=\frac{\rho k^{2}}{4\pi}\frac{\partial}{\partial v}\int_{0}^{t}d\tau  e^{-\tau^{2}/(2\beta)} \left\lbrack \cos(v\tau)\frac{\partial}{\partial v}+\tau\sin(v\tau)\right\rbrack  {P}({v},t-\tau).
\label{m10}
\end{eqnarray}
We note that for the HMF model, the equality (\ref{a3}) is not
satisfied. 

The previous equations assume that the distribution of the bath is
maxwellian.  More generally, if we come back to the
non-Markovian equation Eq.  (\ref{mg2}) and perform the integration on
${\bf r}_1$, we obtain
\begin{eqnarray}
\frac{\partial P}{\partial t}=m(2\pi)^{d}\frac{\partial}{\partial
v^{\mu}}\int_{0}^{t}d\tau\int d{\bf v}_{1}d{\bf k} k^{\mu}k^{\nu}
\frac{\hat{u}(k)^{2}}{|\epsilon({\bf k},{\bf k}\cdot {\bf v}_{1})|^{2}}\cos({\bf k}\cdot {\bf w}\tau)\biggl
(\frac{\partial}{\partial v^{\nu}}
-\frac{\partial}{\partial v_{1}^{\nu}}\biggr ) {P}({\bf v},t-\tau){f}({\bf v}_{1}).\nonumber\\
\label{m11}
\end{eqnarray}
For the HMF model, this equation reduces to
\begin{eqnarray}
\frac{\partial P}{\partial
t}=\frac{k^{2}}{4\pi}\frac{\partial}{\partial
v}\int_{0}^{t}d\tau\int d{v}_{1}\frac{\cos(w\tau)}{|\epsilon(1,v_{1})|^{2}}\biggl
(\frac{\partial}{\partial v}
-\frac{d}{d v_{1}}\biggr ) {P}({v},t-\tau){f}({v}_{1}).
\label{m13}
\end{eqnarray}
This equation is valid for any steady distribution of the field
particles, not only for the statistical equilibrium state (thermal bath).

\section{Conclusion}
\label{sec_conc}

In this paper, starting from the Klimontovich equation and using a
quasilinear theory, we have developed a kinetic theory for systems
with weak long-range interactions. We have obtained general equations that
take into account spatial inhomogeneity and memory effects. These
peculiarities are specific to systems with unshielded long-range
interactions. However, in order to obtain closed kinetic equations we
have been obliged to neglect some collective effects. These collective
effects can be taken into account for spatially homogeneous systems
with short memory time (with respect to the slow collisional
relaxation time). In that case, we recover the Lenard-Balescu equation
of plasma physics with slight modifications (see Paper II and footnote
$2$). It would be valuable to develop a formalism that takes into
account both spatial inhomogeneity and collective effects. This has
been partly done in Paper III, where we have obtained two coupled
Eqs. (III-6)-(III-7) that are exact at order $O(1/N)$. However, it
seems difficult to go any further without either (i) considering
homogeneous systems or (ii) neglecting collective effects.  In fact, due to the huge timescale separation between the dynamical time
$t_{D}$ and the relaxation time $\ge Nt_{D}$, it could be of interest
to develop a kinetic theory in angle-action variables as considered in
\cite{angle}, where an orbit-averaged-Fokker-Planck equation has been derived
for one-dimensional systems with weak long-range interactions (note
that the formalism developed in the present paper could be used to
better justify the kinetic equation obtained in \cite{angle}). This
could be a future direction of investigation. Another direction of
research would be to investigate in deeper detail the non-markovian
kinetic equations derived in this paper. This will be considered in
future works.

\appendix

\section{First and second moments of the velocity  increment}
\label{sec_fs}

In this Appendix, we calculate the first and second moments $\langle
\Delta v^{\mu}\rangle$ and $\langle \Delta v^{\mu}\Delta v^{\nu}\rangle$ of the
increment of velocity of the test particle directly from the
Hamiltonian equations of motion (I-1). We follow a procedure similar
to that used by Valageas \cite{valageas} in a different context. Since
the calculations are similar, we shall only give the main steps of the
derivation. For simplicity, we assume that all the particles have the
same mass $m$. In order to separate the mean field dynamics from the
discrete effects giving rise to the diffusion and to the friction of the
test particle, we write the Hamiltonian (I-1) as
\begin{eqnarray}
H=m (H_{0}+H_{I}),
\label{fs1}
\end{eqnarray}
where we defined the mean field Hamiltonian $H_{0}$ by
\begin{eqnarray}
H_{0}=\frac{1}{2}\sum_{i}v_{i}^{2}+\sum_{i}\Phi_{0}({\bf r}_{i}),
\label{fs2}
\end{eqnarray}
and the interaction Hamiltonian $H_{I}$ by
\begin{eqnarray}
H_{I}=e^{\omega t}\left\lbrack m\sum_{i<j} u({\bf r}_{i}-{\bf r}_{j})-\sum_{i}\Phi_{0}({\bf r}_{i})\right\rbrack.
\label{fs3}
\end{eqnarray}
In Eqs. (\ref{fs2})-(\ref{fs3}) the mean field potential is given by
\begin{eqnarray}
\Phi_{0}({\bf r})=\int \rho({\bf r}')u(|{\bf r}-{\bf r}'|) d{\bf r}',
\label{fs4}
\end{eqnarray}
where $\rho({\bf r}')$ is the mean field equilibrium spatial density
of the particles. The factor $e^{\omega t}$ has been added for the
computation of perturbative eigenmodes and we shall ultimately let
$\omega\rightarrow 0^{+}$. Thus $H_{0}$ describes the mean field
dynamics whereas $H_{I}$ describes the discrete effects which vanish
in the limit $N\rightarrow +\infty$. Therefore, we consider $H_{I}$ as
a perturbation of $H_{0}$ and we apply a perturbative analysis in
powers of $1/N$. We assume that the system is spatially homogeneous so
that $\Phi_{0}=0$. The  interaction Hamiltonian $H_{I}$ can be rewritten
\begin{eqnarray}
H_{I}=e^{\omega t}m\sum_{i<j}\int\hat{u}({\bf k})e^{i{\bf k}({\bf r}_{i}-{\bf r}_{j})}d{\bf k}.
\label{fs5}
\end{eqnarray}
On the other hand, the equations of motion read
\begin{eqnarray}
\frac{d{\bf v}_{i}}{dt}=-\frac{1}{m}\frac{\partial H}{\partial {\bf r}_{i}}=-\frac{\partial }{\partial {\bf r}_{i}}(H_{0}+H_{I}),\qquad  \frac{d{\bf r}_{i}}{dt}=\frac{1}{m}\frac{\partial H}{\partial {\bf v}_{i}}={\bf v}_{i}.
\label{fs6}
\end{eqnarray}
We write the trajectories $\lbrace {\bf r}(t),{\bf v}(t)\rbrace$ as the perturbative expansions ${\bf r}={\bf r}^{(0)}+{\bf r}^{(1)}+{\bf r}^{(2)}+...$ where ${\bf r}^{(k)}$ is formally of order $k$ over $H_{I}$. At zeroth-order, we simply have
\begin{eqnarray}
\frac{d{\bf v}_{i}^{(0)}}{dt}={\bf 0}, \qquad \frac{d{\bf r}_{i}^{(0)}}{dt}={\bf v}_{i}^{(0)},
\label{fs7}
\end{eqnarray}
which yields the rectilinear  orbits
\begin{eqnarray}
{\bf v}_{i}^{(0)}(t')={\rm Cte}={\bf v}_{i}, \qquad {\bf r}_{i}^{(0)}(t')={\bf v}_{i} (t'-t)+{\bf r}_{i},
\label{fs8}
\end{eqnarray}
where, in the following,  ${\bf r}_{i}$ and ${\bf v}_{i}$ denote the position and the velocity of the particle $i$ at time $t$. At first order, we obtain
\begin{eqnarray}
\frac{d{\bf v}_{i}^{(1)}}{dt}=-\frac{\partial H_{I}}{\partial {\bf r}_{i}}, \qquad  \frac{d{\bf r}_{i}^{(1)}}{dt}={\bf v}_{i}^{(1)},
\label{fs9}
\end{eqnarray}
where we can substitute the zeroth-order orbits in the r.h.s. of these
expressions. This yields
\begin{eqnarray}
\frac{d{\bf v}_{i}^{(1)}}{dt}=-\frac{\partial}{\partial {\bf r}_{i}}e^{\omega t}m\sum_{j<j'}\int \hat{u}(k)e^{i{\bf k}({\bf r}_{j}^{(0)}-{\bf r}_{j'}^{(0)})}d{\bf k}.
\label{fs10}
\end{eqnarray}
Using Eq. (\ref{fs8}) and integrating on time, we obtain
\begin{eqnarray}
{\bf v}_{i}^{(1)}=-\frac{\partial}{\partial {\bf r}_{i}}\int_{-\infty}^{t}dt' e^{\omega t'}m\sum_{j<j'}\int \hat{u}(k)e^{i{\bf k}({\bf v}_{j}-{\bf v}_{j'})(t'-t)}e^{i{\bf k}({\bf r}_{j}-{\bf r}_{j'})}d{\bf k}.
\label{fs11}
\end{eqnarray}
Therefore,
\begin{eqnarray}
{\bf v}_{i}^{(1)}=-\frac{\partial}{\partial {\bf r}_{i}}e^{\omega t}m\sum_{j<j'}\int \hat{u}(k)\frac{e^{i{\bf k}({\bf r}_{j}-{\bf r}_{j'})}}{\omega+i{\bf k}({\bf v}_{j}-{\bf v}_{j'})}d{\bf k}.
\label{fs12}
\end{eqnarray}
Substituting this expression in Eq. (\ref{fs9})-b, we get
\begin{eqnarray}
\frac{d{\bf r}_{i}^{(1)}}{dt}=-e^{\omega t}m\sum_{j\neq i}\int \hat{u}(k)\frac{i{\bf k}e^{i{\bf k}({\bf r}_{i}^{(0)}-{\bf r}_{j}^{(0)})}}{\omega+i{\bf k}({\bf v}_{i}^{(0)}-{\bf v}_{j}^{(0)})}d{\bf k}.
\label{fs13}
\end{eqnarray}
Using  Eq. (\ref{fs8}) and integrating on time again, we obtain
\begin{eqnarray}
{\bf r}_{i}^{(1)}=-e^{\omega t}m\sum_{j\neq i}\int \hat{u}(k)\frac{i{\bf k}e^{i{\bf k}({\bf r}_{i}-{\bf r}_{j})}}{\lbrack \omega+i{\bf k}({\bf v}_{i}-{\bf v}_{j})\rbrack^{2}}d{\bf k}.
\label{fs14}
\end{eqnarray}
This can be rewritten
\begin{eqnarray}
{\bf r}_{i}^{(1)}=\frac{\partial}{\partial {\bf v}_{i}}e^{\omega t}m\sum_{j<j'}\int \hat{u}(k)\frac{e^{i{\bf k}({\bf r}_{j}-{\bf r}_{j'})}}{ \omega+i{\bf k}({\bf v}_{j}-{\bf v}_{j'})}d{\bf k}.
\label{fs14b}
\end{eqnarray}
Therefore, at first order, we have
\begin{eqnarray}
{\bf v}_{i}^{(1)}=-\frac{\partial \chi}{\partial {\bf r}_{i}},\qquad {\bf r}_{i}^{(1)}=\frac{\partial \chi}{\partial {\bf v}_{i}},
\label{fs15}
\end{eqnarray}
with
\begin{eqnarray}
\chi=e^{\omega t}m\sum_{j<j'}\int \hat{u}(k)\frac{e^{i{\bf k}({\bf r}_{j}-{\bf r}_{j'})}}{ \omega+i{\bf k}({\bf v}_{j}-{\bf v}_{j'})}d{\bf k}.
\label{fs16}
\end{eqnarray}
At second order, we have
\begin{eqnarray}
\frac{dv_{i}^{\mu(2)}}{dt}=-\sum_{j}\frac{\partial^{2} H_{I}}{\partial r_{i}^{\mu}\partial r_{j}^{\nu}}r_{j}^{\nu(1)}.
\label{fs17}
\end{eqnarray}
The average acceleration of the test particle is
\begin{eqnarray}
\langle \dot{v}^{\mu(2)}\rangle=-\left\langle \frac{\partial^{2} H_{I}}{\partial r^{\mu}\partial r^{\nu}}r^{\nu(1)}\right\rangle-N\left\langle \frac{\partial^{2} H_{I}}{\partial r^{\mu}\partial r_{1}^{\nu}}r_{1}^{\nu(1)}\right\rangle.
\label{fs18}
\end{eqnarray}
Using Eq.  (\ref{fs14b}), the relations
\begin{eqnarray}
\frac{\partial^{2} H_{I}}{\partial r^{\mu}\partial r^{\nu}}=-e^{\omega t}m\sum_{j\neq 0}\int\hat{u}({\bf k})k^{\mu}k^{\nu}e^{i{\bf k}({\bf r}-{\bf r}_{j})}d{\bf k},
\label{fs19}
\end{eqnarray}
\begin{eqnarray}
\frac{\partial^{2} H_{I}}{\partial r^{\mu}\partial r_{1}^{\nu}}=e^{\omega t}m\int\hat{u}({\bf k})k^{\mu}k^{\nu}e^{i{\bf k}({\bf r}-{\bf r}_{1})}d{\bf k},
\label{fs20}
\end{eqnarray}
and performing the averages in Eq. (\ref{fs18}) with respect to the
distribution function $f({\bf v})$, we obtain after
some calculations
\begin{eqnarray}
\langle \dot{v}^{\mu(2)}\rangle=(2\pi)^{d}m e^{2\omega t}\frac{\partial}{\partial v^{\nu}}\int d{\bf v}_{1}d{\bf k}\hat{u}(k)^{2}k^{\mu}k^{\nu}\frac{\omega}{\omega^{2}+({\bf k}\cdot {\bf w})^{2}}f({\bf v}_{1})\nonumber\\
-(2\pi)^{d}m e^{2\omega t}\int d{\bf v}_{1}d{\bf k} f({\bf v}_{1})\frac{\partial}{\partial v^{\nu}}\hat{u}(k)^{2}k^{\mu}k^{\nu}\frac{\omega}{\omega^{2}+({\bf k}\cdot {\bf w})^{2}}.
\label{fs21}
\end{eqnarray}
We introduce the velocity increment $\Delta v^{\mu}=v^{\mu}(t+\Delta
t)-v^{\mu}(t)$. Noting that Eq. (\ref{fs21}) represents the variation
of the velocity increment at order $1/N$,
taking the limit $\omega\rightarrow 0^{+}$ and using ${\rm
lim}_{\omega\rightarrow 0}\omega/(\omega^{2}+x^{2})=\pi\delta(x)$, we
find that
\begin{eqnarray}
\left \langle \frac{\Delta v^{\mu}}{\Delta t}\right\rangle=\pi (2\pi)^{d} m \int d{\bf v}_{1}d{\bf k} f({\bf v}_{1}) \hat{u}(k)^{2}k^{\mu}k^{\nu}\left (\frac{\partial}{\partial v^{\nu}}-\frac{\partial}{\partial v_{1}^{\nu}}\right )\delta({\bf k}\cdot {\bf w}).
\label{fs22}
\end{eqnarray}
On the other hand, the diffusion tensor $\langle \Delta
v^{\mu}\Delta v^{\nu}\rangle$ at order $1/N$ is equal to $\langle
\Delta v^{\mu(1)}\Delta v^{\nu(1)}\rangle$. According to Eq.  (\ref{fs12}), we
have
\begin{eqnarray}
{v}^{\mu (1)}=-i e^{\omega t}m\sum_{j\neq 0}\int \hat{u}(k)k^{\mu}\frac{e^{i{\bf k}({\bf r}^{(0)}-{\bf r}^{(0)}_{j})}}{\omega+i{\bf k}({\bf v}^{(0)}-{\bf v}_{j}^{(0)})}d{\bf k}.
\label{fs23}
\end{eqnarray}
Therefore, using Eq. (\ref{fs8}), we obtain
\begin{eqnarray}
\Delta{v}^{\mu (1)}=-i e^{\omega (t+\Delta t)}m\sum_{j\neq 0}\int d{\bf k} \hat{u}(k)k^{\mu}\frac{e^{i{\bf k}({\bf r}-{\bf r}_{j})+i{\bf k}({\bf v}-{\bf v}_{j})\Delta t}}{\omega+i{\bf k}({\bf v}-{\bf v}_{j})}\nonumber\\
+i e^{\omega t}m\sum_{j\neq 0}\int d{\bf k} \hat{u}(k)k^{\mu}\frac{e^{i{\bf k}({\bf r}-{\bf r}_{j})}}{\omega+i{\bf k}({\bf v}-{\bf v}_{j})}.
\label{fs24}
\end{eqnarray}
Taking the average with respect to the distribution function $f({\bf
v})$, we obtain after some calculations
\begin{eqnarray}
\left\langle \frac{\Delta v^{\mu}\Delta v^{\nu}}{\Delta t}\right\rangle=(2\pi)^{d}m e^{2\omega t}\int d{\bf k}d{\bf v}_{1} f({\bf v}_{1}) k^{\mu}k^{\nu}\frac{\hat{u}(k)^{2}}{\omega^{2}+({\bf k}\cdot {\bf w})^{2}}\nonumber\\
\times \frac{1}{\Delta t} \left\lbrack 1+e^{2\omega \Delta t}-2 e^{\omega\Delta t}\cos ({\bf k}\cdot {\bf w}\Delta t)\right\rbrack.
\label{fs25}
\end{eqnarray}
The limit $\omega\rightarrow 0^{+}$ now gives
\begin{eqnarray}
\left\langle \frac{\Delta v^{\mu}\Delta v^{\nu}}{\Delta t}\right\rangle=2(2\pi)^{d}m\int d{\bf k}d{\bf v}_{1} f({\bf v}_{1}) k^{\mu}k^{\nu}\frac{\hat{u}(k)^{2}}{\Delta t({\bf k}\cdot {\bf w})^{2}} \left\lbrack 1-\cos ({\bf k}\cdot {\bf w}\Delta t)\right\rbrack.
\label{fs26}
\end{eqnarray}
Finally, taking $\Delta t\rightarrow +\infty$ and using ${\rm
lim}_{t\rightarrow +\infty}(1-\cos tx)/tx^{2}=\pi\delta(x)$, we
obtain
\begin{eqnarray}
\left\langle \frac{\Delta v^{\mu}\Delta v^{\nu}}{\Delta t}\right\rangle=2\pi(2\pi)^{d}m\int d{\bf k}d{\bf v}_{1} f({\bf v}_{1}) k^{\mu}k^{\nu}\hat{u}(k)^{2}
\delta({\bf k}\cdot {\bf w}).
\label{fs27}
\end{eqnarray}
This relation can also be obtained from the Kubo formula
$\int_{0}^{+\infty}
\dot{v}^{(1)\mu}(t)\dot{v}^{(1)\nu}(t+\Delta t)d(\Delta t)$ using
Eqs. (\ref{fs8}) and (\ref{fs10}). Equations (\ref{fs22}) and
(\ref{fs27}) return the terms of diffusion and friction obtained in
Sec. \ref{sec_test} when collective effects are neglected.

\section{Collective effects}
\label{sec_coll}

When collective effects are taken into account, the force created by a
particle on the others is modified by the influence of a
``polarization cloud''. This effect can be calculated precisely in the
case of a spatially homogeneous medium. In that case, the linearized Klimontovich equations are
\begin{eqnarray}
\frac{\partial\delta f}{\partial t}+{\bf v}\cdot \frac{\partial\delta f}{\partial {\bf r}}-\nabla\delta\Phi\cdot \frac{\partial f}{\partial {\bf v}}=0,
\label{coll1}
\end{eqnarray}
\begin{eqnarray}
\delta\Phi({\bf r},t)=\int u({\bf r}-{\bf r}')\left\lbrack \delta\rho({\bf r}',t)+m\delta({\bf r}'-{\bf r}_{1}-{\bf v}_{1}t)\right\rbrack d{\bf r}',
\label{coll2}
\end{eqnarray}
where $({\bf r}_{1},{\bf v}_{1})$ represent the position and the
velocity of the test particle (here denoted $1$) at time $t=0$, and we
have made the linear trajectory approximation ${\bf r}_{1}(t)={\bf
r}_{1}+{\bf v}_{1}t$ that is valid at leading order. Taking the
Laplace-Fourier transforms of Eqs. (\ref{coll1}) and (\ref{coll2}), we obtain
\begin{eqnarray}
\delta\hat{f}({\bf k},{\bf v},\omega)=\frac{{\bf k}\cdot \frac{\partial f}{\partial {\bf v}}}{{\bf k}\cdot {\bf v}-\omega}\delta\hat\Phi({\bf k},\omega),
\label{coll3}
\end{eqnarray}
\begin{eqnarray}
\delta\hat\Phi({\bf k},\omega)=(2\pi)^{d}\hat{u}(k)\int \delta\hat{f}({\bf k},{\bf v},\omega)\, d{\bf v}+m\ \hat{u}(k)e^{-i{\bf k}\cdot {\bf r}_{1}}\delta({\bf k}\cdot {\bf v}_{1}-\omega).
\label{coll4}
\end{eqnarray}
Substituting Eq. (\ref{coll3}) in Eq. (\ref{coll4}), we find that
\begin{eqnarray}
\delta\hat\Phi({\bf k},\omega)=m\ \frac{\hat{u}(k)}{\epsilon({\bf k},\omega)}e^{-i{\bf k}\cdot {\bf r}_{1}}\delta({\bf k}\cdot {\bf v}_{1}-\omega),
\label{coll5}
\end{eqnarray}
where $\epsilon({\bf k},\omega)$ is the dielectric function
(II-50). Taking the inverse Laplace-Fourier transform of
Eq. (\ref{coll5}), we obtain the effective field created by particle
$1$ on particle $0$ taking into account collective effects:
\begin{eqnarray}
\Phi(1\rightarrow 0,t)=\int m\ \frac{\hat{u}(k)}{\epsilon({\bf k},{\bf k}\cdot {\bf v}_{1})}e^{i{\bf k}\cdot ({\bf r}(t)-{\bf r}_{1}(t))} d{\bf k}.
\label{coll6}
\end{eqnarray}
Finally, the effective force created by particle $1$ on particle $0$ taking into account collective effects is
\begin{eqnarray}
{\bf F}(1\rightarrow 0,t)=-\int i m {\bf k} \frac{\hat{u}(k)}{\epsilon({\bf k},{\bf k}\cdot {\bf v}_{1})}e^{i{\bf k}\cdot ({\bf r}(t)-{\bf r}_{1}(t))} d{\bf k}.
\label{coll7}
\end{eqnarray}

%\newpage


\begin{thebibliography}{10}

\bibitem{dauxois}  {\small  Dynamics and Thermodynamics of Systems with Long Range Interactions, edited by T. Dauxois, S. Ruffo, E. Arimondo and M. Wilkens, Lect. Not. in Phys. {\bf 602}, Springer (2002).}

\bibitem{hb1}  {\small  P.H. Chavanis, Physica A {\bf 361}, 55  (2006) [Paper I].}

\bibitem{hb2} {\small  P.H. Chavanis, Physica A {\bf 361}, 81 (2006) 
[Paper II].}

\bibitem{hb3} {\small  P.H. Chavanis, preprint [{\tt arXiv:0705.4405}] [Paper III].}

\bibitem{saslaw}  {\small W.C. Saslaw, {\it Gravitational Physics of Stellar and Galactic Systems} (Cambridge Univ. Press, 1985).}

\bibitem{paddy}  {\small T. Padmanabhan, Phys. Rep. {\bf 188}, 285 (1990). }

\bibitem{review}  {\small P.H. Chavanis, Int J. Mod. Phys. B {\bf 20}, 3113 (2006). }

\bibitem{sommeria}  {\small J. Sommeria, {\it Two-Dimensional Turbulence} in:	New trends in turbulence, edited by M. Lesieur, A. Yaglom, F. David, Les Houches Summer School {\bf  74}, 385 (2001). }

\bibitem{tabeling}  {\small P. Tabeling, Phys. Rep. {\bf 362}, 1 (2002). }

\bibitem{houches}  {\small P.H. Chavanis, {\it Statistical mechanics of two-dimensional vortices and stellar systems} in \cite{dauxois}; See also [{\tt cond-mat/0212223}].}

\bibitem{balescubook}  {\small R. Balescu, {\it Statistical Mechanics of Charges Particles} (Interscience, New York, 1963).}

\bibitem{pitaevskii}  {\small E.M. Lifshitz, L.P. Pitaevskii, {\it Physical Kinetics} (Pergamon Press, Oxford, 1981).}

\bibitem{hmf}  {\small T. Dauxois, V. Latora, A. Rapisarda, S. Ruffo, A. Torcini, {\it The Hamiltonian Mean Field Model: from Dynamics to Statistical Mechanics and back} in \cite{dauxois}; See also [{\tt cond-mat/0208456}].}




\bibitem{thesis}  {\small P.H. Chavanis, {\it Contributions \`a la m\'ecanique statistique des tourbillons bidimensionnels. Analogie avec la relaxation violente des syst\`emes stellaires}, Ph.D. thesis, ENS Lyon (1996).}



\bibitem{vortex} {\small  P.H. Chavanis, {\tt [arXiv:0704.3953]}.}

\bibitem{landaud}  {\small  P.H. Chavanis,  Eur. Phys. J. B {\bf 52}, 61
(2006). }

\bibitem{kp}  {\small B.B. Kadomtsev, O.P. Pogutse, Phys. Rev. Lett.  {\bf 25}, 1155 (1970). }

\bibitem{sl}  {\small G. Severne and M. Luwel, Astr. Space Sci.  {\bf 72}, 293 (1980).}

\bibitem{dub}  {\small P.H. Chavanis, {\it Statistical mechanics of violent relaxation in stellar systems}, in: Multiscale Problems in Science and Technology  edited by N. Antonic, C.J. van Duijn, W. Jager and A. Mikelic (Springer, Berlin 2002) {\tt [astro-ph/0212205]}. }

\bibitem{lb}  {\small  D. Lynden-Bell, MNRAS  {\bf 136}, 101 (1967).}

\bibitem{super} {\small  P.H. Chavanis, Physica A {\bf 359}, 177 (2006).}

\bibitem{next05}  {\small  P.H. Chavanis, Physica A {\bf 365}, 102 (2006). }

\bibitem{kandrup1}  {\small H. Kandrup,  ApJ {\bf 244}, 316 (1981).}

\bibitem{csr}  {\small  P.H. Chavanis, J. Sommeria and R. Robert, ApJ  {\bf 471}, 385 (1996).}

\bibitem{landaugen}  {\small  P.H. Chavanis, Physica A {\bf 332}, 89 (2004).}

\bibitem{hohl}  {\small F. Hohl, J.W. Campbell, Astron. J.  {\bf 73}, 611 (1968) }

\bibitem{rsmepp}  {\small  R. Robert and J. Sommeria, Phys. Rev. Lett. {\bf 69}, 2776 (1992)}

\bibitem{staquet}  {\small  J. Sommeria, C. Staquet, R. Robert, J. Fluid Mech. {\bf 233}, 661 (1991)}

\bibitem{antoniazzi1}  {\small A. Antoniazzi, D. Fanelli, J. Barr\'e, P.H. Chavanis, T. Dauxois, S. Ruffo,  Phys. Rev. E  {\bf 75}, 011112 (2007).}

\bibitem{tsallis}  {\small  C. Tsallis, J. Stat. Phys.  {\bf 52},
479 (1988).}

\bibitem{cstsallis}  {\small  P.H. Chavanis and C. Sire, Physica A  {\bf 356},
419 (2005).}

\bibitem{boghosian}  {\small  B.M. Boghosian, Phys. Rev. E {\bf 53}, 4754 (1996)}

\bibitem{brands}  {\small  H. Brands, P.H. Chavanis, R. Pasmanter, J. Sommeria, Phys. Fluids {\bf 11}, 3465 (1999)}

\bibitem{lrt}  {\small V. Latora, A. Rapisarda, C. Tsallis,  Physica A  {\bf 305}, 129 (2002).}

\bibitem{campa}  {\small A. Campa, A. Giansanti, G. Morelli, [arXiv:0706.3664] }

\bibitem{ts}  {\small A. Taruya, M. Sakagami, Phys. Rev. Lett.  {\bf 90}, 181101 (2003) }

\bibitem{yamaguchi}  {\small Y.Y. Yamaguchi, J. Barr\'e, F. Bouchet, T. Dauxois, S. Ruffo,  Physica A  {\bf 337}, 36 (2004).}


\bibitem{plr}  {\small A. Pluchino, V. Latora, A. Rapisarda,  Physica D  {\bf 193}, 315 (2004).}

\bibitem{tlb}  {\small  P.H. Chavanis,  Eur. Phys. J. B {\bf 53}, 487 (2006). }

\bibitem{bt}  {\small J. Binney, S. Tremaine, {\it Galactic Dynamics} (Princeton Series in Astrophysics, Princeton, 1987).}


\bibitem{mineau}  {\small P. Mineau, M.R. Feix, J.L. Rouet, Astron. Astrophys.  {\bf 228}, 344 (1990) }

\bibitem{mk}  {\small H. Morita, K. Kaneko, Phys. Rev. Lett.  {\bf 96}, 050602 (2006) }




\bibitem{bd}  {\small F. Bouchet, T. Dauxois, Phys. Rev. E {\bf 72}, 5103 (2005).}

\bibitem{curious}  {\small  P.H. Chavanis,  Eur. Phys. J. B {\bf 52}, 47
(2006).   }

\bibitem{ct}  {\small S. Chandrasekhar,  ApJ {\bf 99}, 47 (1944).}

\bibitem{kandrup2}  {\small H. Kandrup,  Astro. Space. Sci. {\bf 97}, 435 (1983).}

\bibitem{hubbard}  {\small J. Hubbard, Proc. Roy. Soc. (London) A {\bf 260}, 114 (1961). }

\bibitem{kc}  {\small S. Chandrasekhar,  ApJ {\bf 97}, 255 (1943).}

\bibitem{lenard}  {\small A. Lenard, Ann. Phys. (N.Y.)  {\bf 10}, 390 (1960).}

\bibitem{balescu}  {\small R. Balescu, Phys. Fluids {\bf 3}, 52 (1960).}

\bibitem{bouchet}  {\small F. Bouchet, Phys. Rev. E {\bf 70}, 036113 (2004). }

\bibitem{cvb}  {\small P.H. Chavanis, J. Vatteville, F. Bouchet,   Eur. Phys. J. B {\bf 46}, 61 (2005). }

\bibitem{angle}  {\small  P.H. Chavanis, Physica A {\bf 377}, 469 (2007). }

\bibitem{valageas}  {\small  P. Valageas, Physical Review E {\bf 74}, 016606 (2006).  }







\end{thebibliography}
\end{document}